\documentclass[journal]{IEEEtran}
\usepackage{verbatim}
\usepackage{subfiles}
\usepackage{cite}
\usepackage[justification=centering]{caption}
\usepackage{amsmath}
\usepackage{amssymb}
\usepackage{lipsum}
\usepackage{graphicx}
\usepackage{subcaption}
\usepackage{changepage}
\usepackage{adjustbox}
\usepackage{arydshln}
\usepackage{threeparttable}

\usepackage{array}     
\usepackage{booktabs}
\usepackage{multirow}
\usepackage{siunitx} 
\usepackage{bm}
\usepackage{makecell} 
\usepackage{nomencl}
\usepackage{verbatim}
\usepackage{enumitem}
\usepackage{float}
\usepackage{booktabs}

\makenomenclature

\usepackage{etoolbox}

\usepackage{upgreek}

\usepackage{mathtools}

\usepackage{xcolor, soul} 

\usepackage{ifthen}
 \renewcommand{\nomgroup}[1]{%
  \item[\bfseries
  \ifthenelse{\equal{#1}{A}}{Parameters}{%
  \ifthenelse{\equal{#1}{B}}{Variables}{%
  \ifthenelse{\equal{#1}{O}}{Indices and sets}{}}}%
  ]}

\ifCLASSINFOpdf
\else
\fi
\begin{document}
%
\title{LLM-Enhanced Multi-Agent Reinforcement Learning with Expert Workflow for Real-Time P2P Energy Trading}
%
%
%

\author{Chengwei Lou, Zekai Jin, Wei Tang, Guangfei Geng, Jin Yang,~\IEEEmembership{\textit{Senior Member,~IEEE},}
        Lu Zhang,~\IEEEmembership{\textit{Senior Member,~IEEE}}

\thanks{This work was supported by Smart Grid-National Science and Technology Major Project(2024ZD0800500).  (Corresponding authors: Zekai Jin; Lu Zhang). Chengwei Lou, Zekai Jin,  Wei Tang, and Lu Zhang are with the College of Information and Electrical Engineering, China Agricultural University, Beijing, China. Guangfei Geng is with the School of Information Science and Technology, Guangdong University of Foreign Studies, Guangzhou, China. Jin Yang is with the James Watt School of Engineering, University of Glasgow, Glasgow, United Kingdom. }

}

%
%

\markboth{\parbox{\textwidth}{This manuscript has been accepted for publication in IEEE Transactions on Smart Grid. \\ Copyright has been transferred to IEEE. Reuse of this material is subject to IEEE copyright restrictions}}%
{Shell \MakeLowercase{\textit{et al.}}: Bare Demo of IEEEtran.cls for IEEE Journals}
%



\maketitle

\sethlcolor{white}

\begin{abstract}
Real-time peer-to-peer (P2P) electricity markets dynamically adapt to fluctuations in renewable energy and variations in demand, maximizing economic benefits through instantaneous price responses while enhancing grid flexibility. However, scaling expert guidance for massive personalized prosumers poses critical challenges, including diverse decision-making demands and a lack of customized modeling frameworks. This paper proposes an integrated large language model-multi-agent reinforcement learning  (LLM-MARL) framework for real-time P2P energy trading to address challenges such as the limited technical capability of prosumers, the lack of expert experience, and security issues of distribution networks. LLMs are introduced as experts to generate personalized strategies, guiding MARL under the centralized training with decentralized execution (CTDE) paradigm through imitation. \hl{To handle the scalability issues inherent in large-scale P2P networks, a differential attention-based critic network is introduced to efficiently extract key interaction features and enhance convergence.} Experimental results demonstrate that LLM-generated strategies effectively substitute human experts. The proposed imitative expert MARL algorithms achieve significantly lower economic costs and voltage violation rates on test sets compared to baseline algorithms, while maintaining robust stability. This paper provides an effective solution for the real-time decision-making of the P2P electricity market by bridging expert knowledge with agent learning.
\end{abstract}

\begin{IEEEkeywords}
P2P Energy Trading, Large Language Model, Multi-Agent Reinforcement Learning, Imitation Learning, Attention Mechanism
\end{IEEEkeywords}

%
\IEEEpeerreviewmaketitle

%
%
%
%

\section{INTRODUCTION}

\IEEEPARstart{T}{he} rise of peer-to-peer (P2P) energy trading has shifted electricity users from traditional consumers to "prosumers," combining both production and consumption \cite{CAPPER2022112403}. However, this development faces two challenges: on the virtual layer, prosumers often lack the technical capability for repeated trading and efficient energy management \cite{FENG2025125283}; on the physical level, ensuring system security during the transmission of electricity transactions from the virtual layer in actual distribution networks remains a challenge \cite{9744103}. \hl{Although the majority of energy trading is settled in the day-ahead stage, actual load and renewable generation often deviate substantially from the scheduled profiles due to inherent uncertainties. Consequently, real-time mechanisms are indispensable for rapidly rescheduling trades to restore power balance and enabling distributed resources to provide immediate balancing services}\cite{9416277}. \hl{At present, electricity markets such as Nord Pool in the Nordic region, PJM and ERCOT in the United States, and Australia’s National Electricity Market  mainly operate under a two-stage framework with separate settlement in the day-ahead and real-time markets. As the real-time market is a critical component of electricity market operations, it is necessary to operate P2P market mechanisms in a real-time operational mode as well}\cite{9536422}.

\hl{In response to this requirement, recent studies have focused on developing real-time P2P energy trading. Specifically, studies such as} \cite{LIU2023120216} \hl{and} \cite{ZHENG2025114530} \hl{employed Lyapunov optimization to relax temporal coupling constraints and utilized the Alternating Direction Method of Multipliers to achieve distributed solutions for real-time P2P energy rescheduling. In parallel,} \cite{WANG2024109547} \hl{proposed a real-time P2P energy trading method grounded in Model Predictive Control (MPC). Nevertheless, these approaches have inherent limitations. Lyapunov optimization inevitably introduces approximation errors due to the relaxation of temporal coupling constraints. Similarly, the open-loop nature of MPC policies renders them intrinsically suboptimal in the presence of stochastic noise }\cite{4717266}. \hl{Furthermore, such conventional model-driven methods rely heavily on precise system modeling and parameterization, which are typically difficult to obtain in practice} \cite{9397299}. \hl{Consequently, they often struggle to accommodate the highly dynamic nature of real-time P2P energy trading, creating a trade-off between computational efficiency and decision-making flexibility} \cite{XU2024123923}. \hl{To address these challenges, reinforcement learning (RL) has emerged as a promising solution, owing to its robust adaptability to uncertain environments and its ability to facilitate rapid strategy generation} \cite{11074719}.

\hl{RL relies heavily on trial-and-error interactions with the environment, faces fundamental challenges when applied to P2P energy trading systems with strict operational constraints. RL methods struggle to efficiently converge to a global market equilibrium} \cite{paine2019makingefficientusedemonstrations}, \hl{particularly in large-scale P2P markets characterized by highly personalized prosumers, whose preferences and capacities demand individualized and incentive-compatible decision-making rather than uniform policies} \cite{10946681}. To mitigate these challenges, optimization-based solvers and imitation learning techniques have been introduced to incorporate expert knowledge into the learning process, thereby improving training efficiency and solution feasibility in related power system applications, such as microgrid operation \cite{9585298} and distribution network reconfiguration \cite{9424985}. \hl{For instance,}  \cite{10689197} \hl{proposed a multi-virtual-agent imitation learning framework that leverages adversarial imitation across multiple virtual environments to achieve robust microgrid energy scheduling under uncertain power supply interruptions.} \hl{However, imitation learning approaches driven by a single expert’s decision-making experience remain limited in their ability to adapt to the diverse personalities and strategic preferences of prosumers in P2P energy trading. Standardized expert strategies often fail to capture individualized behaviors and incentive structures. Moreover, relying on a human Distribution System Operator (DSO) as the expert decision-maker introduces practical constraints, including high labor costs and limited real-time responsiveness, while the lack of generalization capability further amplifies operational risks} \cite{huang2025orlmcustomizableframeworktraining}.

With the emergence of large language models (LLMs) like ChatGPT, LLMs showcase strong reasoning, decision-making, and generalization abilities, addressing the shortcomings of using single experts in RL to handle the heterogeneity of prosumers, as well as the challenges of human expert labor costs and generalization. \hl{While LLMs have begun to be applied in the power and energy sector—such as in} \cite{11016112}, \hl{where an $RL^2$ mechanism is designed with LLM-assisted safe reinforcement learning for active distribution network energy management via an iterative penalty function, and in} \cite{10339881}, \hl{where the LLM is used to interpret linguistic operational stipulations and generate reward signals, alleviating the need for manually designing explicit reward functions for qualitative objectives.} \cite{10688670} evaluates LLMs' performance in various power system tasks, highlighting their potential in complex system modeling and reasoning. The study in \cite{jia2025enhancingllmspowersimulations} introduces a power multi-agent framework with a feedback mechanism, validated on the DALINE and MATPOWER platforms. However, the integration of LLMs as expert systems for assisting prosumers in RL training for P2P energy trading remains underexplored. Currently, in non-power system domains LLMs have been successfully used as expert guides in autonomous driving RL training \cite{xu2025telldriveenhancingautonomousdriving, pang2024largelanguagemodelguided}, providing valuable insights for the application of LLMs in the P2P energy trading domain.


\hl{Nevertheless, applying single-agent RL to P2P energy trading entails inherent limitations. When consumers seek individual economic optimality, uncoordinated decisions can cause an excessive concentration of load or generation, posing risks to the security of the distribution network} \cite{7122924}. \hl{Furthermore, the concurrent and independent policy updates of multiple prosumers render the learning environment highly non-stationary from the perspective of any single agent} \cite{DBLP:journals/corr/Hernandez-LealK17}. To address these challenges, multi-agent reinforcement learning (MARL) has emerged as a promising paradigm for P2P energy trading, as it explicitly captures strategic interactions among distributed prosumers in continuous action spaces, supports decentralized decision-making, and promotes cooperative behaviors. \cite{10444990} proposed a multi-agent adversarial reinforcement learning to solve the active voltage control problem in peer-to-peer energy trading-enabled distribution networks. \cite{10966432} formulates the network-constrained MARL P2P energy trading problem as a cooperative Markov game.

However, the complexity of managing numerous agents with high-dimensional actions under the Centralized Training and Decentralized Execution (CTDE) framework\cite{KRAEMER201682} limits the ability to exploit global collaborative information \cite{zhou2025centralizedtrainingdecentralizedexecution}. To improve learning efficiency, attention mechanisms have been introduced into MARL models \cite{9283339} to better extract relevant features. Empirical studies confirm their effectiveness in power system applications such as voltage regulation \cite{10587051}, microgrid trading \cite{9960828}, and community-based P2P trading \cite{9509579}. \hl{While attention is inherently a neural network architecture rather than a market mechanism, it is particularly well-suited for modeling the uneven interaction patterns in P2P energy trading}\cite{10571970}. \hl{Nevertheless, conventional attention mechanisms often lacks a holistic understanding of the entire sequence and may ignore critical information}  \cite{SAVINO2025125993}. \hl{In a distribution network, a prosumer’s state is significantly affected only by a subset of relevant peers, rather than the entire population. Standard attention mechanisms may fail to distinguish these critical trading partners from irrelevant ones, leading to inefficient coordination in large-scale P2P systems, which constrains MARL's performance in real-world P2P trading.}


\hl{In summary, this paper addresses the P2P energy trading limitations of existing reinforcement learning methods identified in TABLE} \ref{tab:innovation_comparison}. \hl{It proposes a novel integrated framework in which LLM-based experts guide the training of MARL agents, with the goal of maximizing social welfare in local real-time P2P electricity markets featuring prosumer collaboration. During the initial stage of training, the LLM generates model code tailored to differentiated prosumers’ demands. Specifically, each prosumer’s expert is embedded within a multi-LLM workflow, where real-time states are fed into solvers, indirectly generating expert strategies for each prosumer. For the training phase, a CTDE-based imitative expert MARL algorithm is proposed. Furthermore, inspired by the recent application of the Differential Transformer in LLMs} \cite{ye2025differentialtransformer}, \hl{an enhanced critic network architecture utilizing differential attention is designed. This architecture is specifically developed to mitigate irrelevant information interference from other agents among large-scale prosumers during training and to enhance the overall convergence performance of the algorithm. The main contributions of this article are summarized as follows:}

\begin{table}[htbp]
  \centering
  \caption{Comparison limitations in P2P energy trading reference}
  \label{tab:innovation_comparison}
  \setlength{\tabcolsep}{5pt}
  \begin{tabular}{lcccc}
    \toprule
    \textbf{Ref.} & \makecell[c]{\textbf{Prosumer} \\ \textbf{Personalized}} & \makecell[c]{\textbf{Network} \\ \textbf{Constraints}} & \makecell[c]{\textbf{Large} \\ \textbf{Scalability}} & \textbf{Method} \\
    \midrule
    \cite{10946681}  & $\checkmark$ & $\times$ & \textemdash & RL\\
    \cite{10966432}  & $\times$ & $\checkmark$ & $\times$ & MARL\\
    \cite{9509579}  & $\times$ & $\times$ & $\checkmark$ & MARL\\
    \cite{10571970}  & $\times$ & $\times$ & $\checkmark$ & MARL\\
    \midrule
    This Paper & $\checkmark$ & $\checkmark$ & $\checkmark$ & LLM-MARL\\
    \bottomrule
  \end{tabular}
\end{table}



\begin{enumerate}[label=\arabic*)]
\item A novel LLM-MARL integrated framework is proposed for the real-time P2P electricity market. By introducing LLMs as expert in the P2P energy trading, the framework replaces human experts to guide MARL agents during training. \hl{This directly substitutes the process of human expert code generation or expert guidance and achieves a deep integration of expert knowledge and LLM-based reasoning.}
\item \hl{An LLM expert workflow tailored to local P2P energy trading is developed for each prosumer. This workflow transforms prosumer natural language input into executable actions through model generation, tool retrieval, code generation, and code correction. By processing state information, it dynamically generates prosumer strategies that balance economic performance and distribution network security, thereby providing reliable expert guidance during training.}
\item A novel imitative expert MARL algorithm is proposed. It introduces the Wasserstein metric to measure the similarity between expert strategy and agent policy, enabling effective guidance from the LLM experts' workflow. Furthermore, a differential multi-head attention-based Critic network is designed to improve policy evaluation accuracy and accelerate the learning process in the large-scale P2P energy trading, thereby boosting overall algorithmic performance.
\end{enumerate}

This paper is structured as follows: Section 2 introduces the system model and decentralized partially observable Markov decision process (Dec-POMDP) formulation for P2P energy trading. Section 3 details the integration of LLM and MARL. Section 4 presents numerical study and result analysis. Section 5 concludes the paper with key findings.
\section{PRELIMINARIES}
\subsection{Prosumer energy management and P2P Trading}


Prosumers are modeled as independent energy units with conventional distributed generators (CDGs), renewable distributed generators (RDGs), battery energy storage systems (BESSs), and controllable loads (CLs). This study adopts a 15-minute interval for optimization to align with real-time P2P energy trading settlement practices\cite{LIU2023120216}. 


CDG output is limited by physical and safety constraints, including ramp rate limits on power variation. RDGs, such as Photovoltaic (PV) and Wind Turbine (WT), adjust active and reactive power via inverter control. BESS regulates energy flow within power and State of Charge (SOC) limits to prevent overcharging or deep discharge. CLs offer demand-side flexibility, adjusting consumption based on load characteristics and user preferences.
\begin{align}
&    P_{i,\mathrm{\min}}^{CDG}\leq P_{i,t}^{CDG}\leq P_{i,\mathrm{\max}}^{CDG},\\ 
&   |P_{i,t}^{CDG}-P_{i,t-1}^{CDG}|\leq R_{i,\mathrm {m a x}}^{CDG},\\
&0 \leq P_{i,t}^{RDG}\leq P_{i,t,\max}^{RDG},  \\
&P_{i,t}^{RDG,2}+Q_{i,t}^{RDG,2}\le S_{i,\max}^{RDG,2},  \\
&    P_{i,\min}^{BESS}\leq P_{i,t}^{BESS}\leq P_{i,m a x}^{BESS},\\
 &   S O C_{i,m i n}^{B E S S}\leq S O C_{i,t}^{BESS}\leq S O C_{i,m a x}^{B E S S S},\\
&SOC_{i,t}^{BESS}=\begin{small}\left\{ \begin{array}{*{35}{l}}
   SOC_{i,t-1}^{BESS}+P_{i,t}^{BESS}/\eta ,P_{i,t}^{BESS}<0  \\
   SOC_{i,t-1}^{BESS}+\eta P_{i,t}^{BESS},P_{i,t}^{BESS}\ge 0 
\end{array} \right.\end{small}\\
 &   0\leq P_{i,t}^{CL}\leq \alpha P_{i,t}^{Load},
\end{align}
where $P_{i,t}^{CDG}$ is the CDG output, bounded by $[P_{i,\mathrm{min}}^{C D G},P_{i,\mathrm{max}}^{C D G}]$ with ramp limit $R_{i,\mathrm{max}}^{CDG}$; $P_{i,t}^{RDG}$,$Q_{i,t}^{RDG}$ is the active / reactive power of RDG, limited by its maximum active power $P_{i,t, \mathrm{max}}^{RDG}$ and apparent power rating $S_{i, \mathrm{max}}^{RDG}$, $P_{i,t}^{BESS}$ is the power of BESS bounded by $[P_{i,\mathrm{min}}^{BESS},P_{i,\mathrm{max}}^{BESS}]$; $SOC_{i,t}^{BESS}$ obeys efficiency $\eta$ and bounded by $[SOC_{i,\mathrm{min}}^{BESS},SOC_{i,\mathrm{max}}^{BESS}]$; $P_{i,t}^{CL}$ is the controllable load up to fraction $\alpha$ of its demand.

Prosumers at different nodes engage in P2P energy trading to increase their revenue. Each prosumer must satisfy an internal power balance, ensuring that generation, consumption, and storage remain aligned.
\begin{align}
 & \begin{aligned}
  &P_{i,t}^{EX} =-P_{i,t}^{Grid} -P_{i,t}^{P2P} \\
  & \textcolor{white}{P_{i,t}}       = P_{i,t}^{CDC} + P_{i,t}^{RDG} + P_{i,t}^{CL} - P_{i,t}^{Load} - P_{i,t}^{BESS},\\
  \end{aligned}\\
   & Q_{i,t}^{EX}  = Q_{i,t}^{RDG} - Q_{i,t}^{Load},
    \end{align}
where $P_{i,t}^{P2P}$ is the P2P electricity trading; $P_{i,t}^{Grid}$ is the active power purchase and sale with the grid; $P_{i,t}^{Load}$,$Q_{i,t}^{Load}$ are the active and reactive loads of prosumer.

Power flow constraints in the distribution network guarantee the safe, stable, and efficient operation of the power system.
\begin{align}
    &P_{i,t}^{EX}=V_{i,t}\sum_{j\in{\cal N}}V_{j,t}(G_{i j}\cos\theta_{i j,t}+B_{i j}\sin\theta_{i j,t}),\\
    & Q_{i,t}^{EX}=V_{i,t}\sum_{j\in{\cal N}}V_{j,t}(-B_{i j}\cos\theta_{i j,t}+G_{i j}\sin\theta_{i j,t}),\\
    & V_{m i n}\leq V_{i,t}\leq V_{m a x},
 \end{align}    
where $V_{i,t}$ is the node voltage magnitude and bounded by $[V_{\mathrm{min}},V_{\mathrm{max}}]$; $G_{ij}$,$B_{ij}$ is the conductance and susceptance of branch ij; $\theta_{ij}$ is the voltage phase angle difference.

The total operational cost for a single prosumer$C_{i}^{Cost}$ consists of the power purchase or sale cost from the grid $C_{i}^{Grid}$, CDG operational costs $C_{i}^{CDG}$, BESS maintenance costs $C_{i}^{BESS}$, CL compensation costs $C_{i}^{CL}$, and P2P energy trading costs $C_{i}^{P2P}$.
\begin{equation}
    C_{i}^{Cost} = C_{i}^{Grid} + C_{i}^{CDC} + C_{i}^{BESS} + C_{i}^{CL} + C_{i}^{P2P}.
\end{equation}

Prosumers’ electricity purchase and sale costs follow time-of-use pricing. CDG operational costs are modeled as quadratic functions of fuel consumption. BESS costs depend on charging and discharging power, while CL costs reflect user dissatisfaction from load reduction. P2P trading incurs additional trading costs.
\begin{align}
	&    C_{i}^{Grid} =\sum_{t} \left\{ \begin{array}{*{35}{l}}
		\lambda_{t}^{S} P_{i,t}^{Grid} ,P_{i,t}^{Grid}<0  \\
		\lambda_{t}^{B} P_{i,t}^{Grid},P_{i,t}^{Grid}\ge 0  \\
	\end{array} \right.\\
	&    C_{i}^{CDG}=\sum_{t} c^{CDG} P_{i,t}^{CDG,2}+b^{CDG} P_{i,t}^{CDG},\\
	&    C_{i}^{BESS}=\sum_{t}\gamma |P_{i,t}^{BESS}|,\\
	&    C_{i}^{CL}=\sum_{t} \rho |P_{i,t}^{CL}|,\\
	&    C_{i}^{P2P} = \sum_{t}\lambda^{DSO}|P_{i,t}^{P2P}| + \lambda_{t}^{P2P} P_{i,t}^{P2P},
\end{align}
where $\lambda_{t}^{B}$, $\lambda_{t}^{S}$ are the time-of-use electricity purchase and sales price for the grid; $c^{CDG}$, $b^{CDG}$ are the quadratic and linear cost coefficients of CDG fuel cost; $\gamma$ is the maintenance cost coefficient; $\rho$ is the compensation cost coefficient; $\lambda_{t}^{DSO}$ is the P2P compensate fees charged by DSO; $\lambda_{t}^{P2P}$ is the real-time P2P price.

{The real-time P2P trading price $\lambda_{t}^{P2P}$ is set as a fixed proportion between the grid’s time-of-use buying and selling prices}\cite{WOS:001154491500001}. {It is expressed as:}
\begin{equation}
	\lambda_{t}^{P2P} = \kappa^{P2P} (\lambda_{t}^{B} - \lambda_{t}^{S}) + \lambda_{t}^{S},
\end{equation}
{where $\kappa^{P2P} \in [0,1]$ is a price coefficient that determines the fairness of the P2P market. This pricing scheme incentivizes local trading by offering sellers a higher price than grid export rates and buyers a lower price than grid retail rates.}

This paper aims to maximize the social welfare of prosumers by assuming that each prosumer makes rational trading decisions and accepts a centrally coordinated P2P price determined by the DSO. Whenever a prosumer experiences a surplus or deficit of electricity, it first seeks to balance supply and demand through local P2P energy trading.
\begin{equation}
	\sum_{i} \sum_{t} \lambda_{t}^{P2P} P_{i,t}^{P2P} = 0,
\end{equation}

However, within P2P energy trading, the total revenue of prosumers is 0\cite{10804212}.

\subsection{Dec-POMDP for P2P Energy Trading}


In P2P energy trading, the inherent uncertainty challenges traditional mathematical optimization methods in meeting the precision and real-time demands of prosumer optimization control. RL methods can address these limitations, enabling data-driven optimization decisions. This paper models the P2P trading problem for multiple prosumers in a distribution network as a Dec-POMDP, represented as an eight-tuple $\langle I, A, S, O, P, r, \pi, \gamma \rangle$, where $S$ is the global state space, $A$ is the joint action space, $r$ is the global reward function based on state transitions $P$, $O$ is the observation space, and $\gamma$ is the discount factor. The model captures the features of information asymmetry and decentralized decision-making through partial observability and decentralized architecture.

    
    

    
    

\begin{enumerate}
    \item Agent:
       Each prosumer participating in P2P energy trading at a node in the distribution network is considered an agent. The set of agents is defined as $I$.
    
    \item Action $A$:
        The joint action space at time $t$ is represented by $a_{t}=\{a_{i,t}\,|i\in{I}\}$, where $\forall a_{t}\in{A}$. The action space of each agent, $a_{i,t}$, consists of the actions of controllable devices within the prosumer.
    \begin{equation}
a_{i,t}=\biggr[P_{i,t}^{CDG},P_{i,t}^{RDG},  Q_{i,t}^{RDG},P_{i,t}^{BESS},P_{i,t}^{CL}\biggl].
    \end{equation}
    
    \item State $S$:
       The global state at time $t$, denoted as $s_{t}=\{s_{i,t}\,|i\in{N}\}$, for $\forall s_{t}\in{N}$, encompasses the operational conditions of all nodes in the distribution network.  $V_{i,t-1}$ \hl{ captures node voltage magnitudes via phasor measurement units and wireless sensors}\cite{11087389}, \hl{enabling realistic prosumer state observation}. Specifically, $s_{i,t}$ includes the operational state of the prosumer, the distribution network's interaction state, and the previous action of the prosumer:
       
    \begin{equation}
        \begin{aligned}
{s}_{i,t} = \biggr[ &t,\lambda_{t}^{B},\lambda_{t}^{S},P_{i,t, \text{max}}^{RDG}, P_{i,t}^{Load}, Q_{i,t}^{Load}\\
& P_{i,t-1}^{CDG},SOC_{i,t-1},P_{i,t-1}^{P2P}, V_{i,t-1} \biggr].
\end{aligned}
    \end{equation} 

    \item Observation $O$:
  
        The joint observation at time $t$ is represented by $o_{t}=\{o_{i,t}\,|i\in{I}\}$, for $\forall o_{t}\in{I}$. The observation of the $i$-th agent at time $t$ is the state of the corresponding node, i.e., $o_{i,t} = s_{i,t}$.
    
    \item State Transition Probability $P$:
       The state transition is described by the conditional probability distribution $P\bigl(s_{t+1}\bigr|s_{t}, a_{t}\bigr)$, which represents the probability of transitioning to the next time step. This transition process considers power flow distribution, load demand fluctuations, and renewable energy output uncertainty. The power flow distribution is driven by the actions $a_{t}$ of the controlled devices.

    \item Reward Function $r$:

     \hl{Given the strong coupling between active power injections and nodal voltages in P2P energy trading, ignoring physical limits renders market outcomes infeasible}\cite{10444990}. \hl{Following} \cite{10966432}, \hl{we adopt a unified global reward function shared among all agents. This design is crucial because, in physically coupled networks, a shared signal is necessary to guide agents toward a Nash Equilibrium that maximizes social welfare while strictly adhering to voltage safety limits. To this end, the reward function is designed as:}

\begin{align}
    &r =  r^{Cost}+r^{Pen},\\
    &r^{Cost} = \delta \sum_{I}  C_{i}^{Cost},\\
    &\scalebox{0.9}{$r^{Pen} = a^{Pen} + c^{Pen}{\mathrm{\max}}\biggl\{0,\biggl|{\ {V}_{base}-{V}_{i,t}}\biggr|-{\frac{{{{V}_{\max}}}-{ {{V}_{\min}}}}{2}}\biggr\}$},
\end{align}
where $\delta$ is the weight coefficient for operational costs, and $r^{\text{Pen}}$ represents the penalty cost for voltage violations in the distribution network. 
\end{enumerate}

\section{METHODOLOGY}

\begin{figure}
	\centering
	\includegraphics[width=0.8\linewidth]{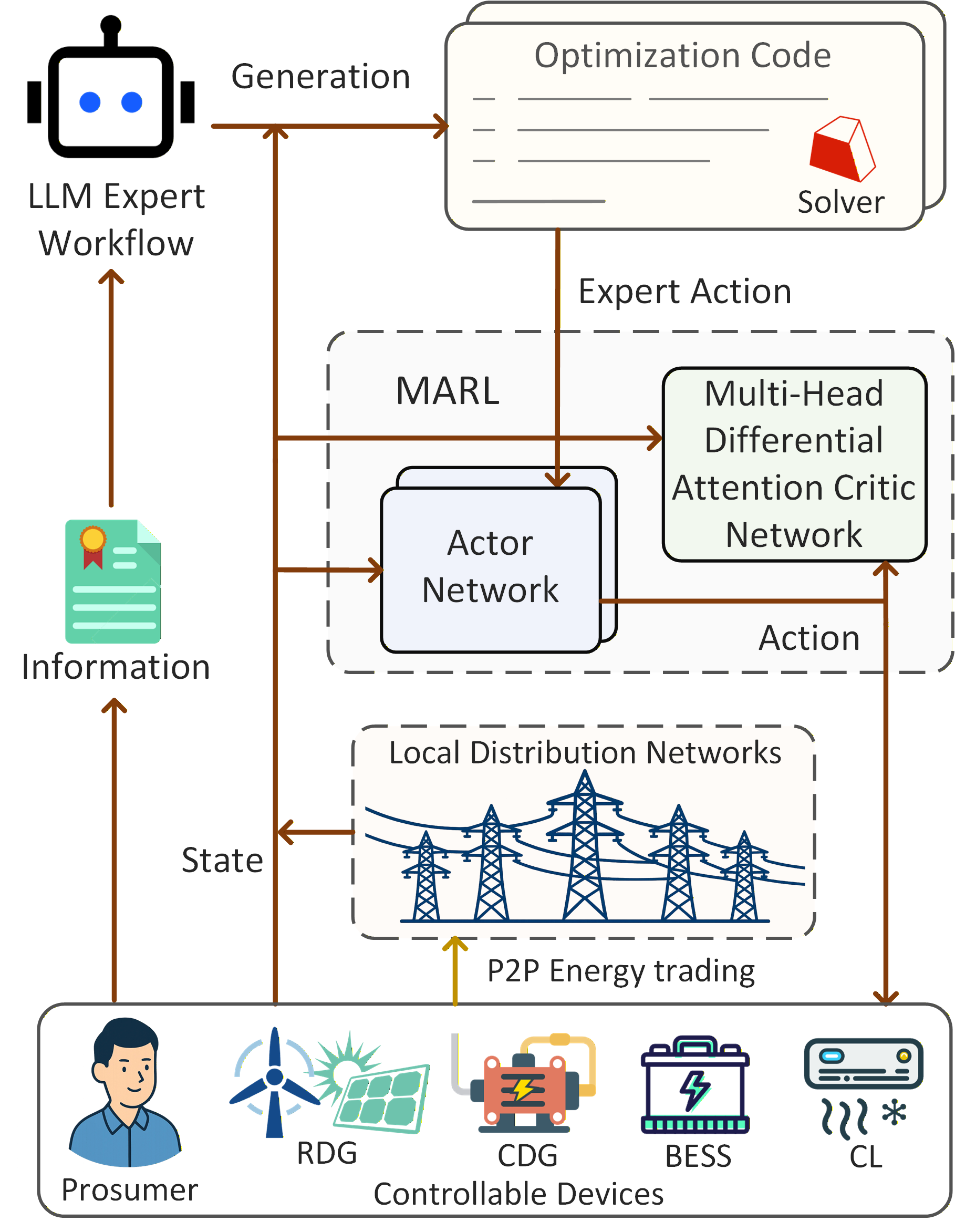}
	\caption{Our proposed LLM-MARL framework}
	\label{fig:framework}
\end{figure}

This paper focuses on P2P energy trading in distribution networks and introduces a novel MARL framework constrained by expert strategies. The proposed framework adopts an off-policy RL paradigm to enable efficient online updates. In this approach, LLM serves as an expert, generating personalized strategies to guide prosumers in energy trading. {Fundamentally, it constitutes a form of “phase-triggered code synthesis training” enabled by LLM. Specifically, at the onset of each training phase, the LLM synthesizes optimization code tailored to the current system configuration. During subsequent online learning iterations, this code is executed in a loop to produce state-dependent, personalized strategies for each prosumer agent.} To ensure a principled integration of expert knowledge with agent learning, the imitative expert MARL algorithm is enhanced using the Wasserstein metric, which promotes alignment between the LLM strategy and the learned agent policies. {When prosumer devices are modified or newly added, the LLM regenerates the code to adapt to the updated environment.} The overall framework is illustrated in Fig. \ref{fig:framework}.


\subsection{LLM Expert Workflow}

\subsubsection{Knowledge Enhancement Methods}
Knowledge injection enhances the capabilities of LLMs and help mitigate information hallucination when these models act as domain experts. The main knowledge enhancement approaches include Supervised Fine-Tuning (SFT) and Retrieval-Augmented Generation (RAG). Unlike SFT, which requires full parameter fine-tuning, RAG decouples the knowledge storage from the inference generation. It only requires maintaining an external knowledge base. This characteristic not only ensures the real-time model outputs but also improves the interpretability of the generated results through  explicit knowledge tracing. Recent research from Microsoft has demonstrated that RAG outperforms SFT in integrating domain-specific knowledge into LLMs \cite{XIAO2024114691}, making RAG the core knowledge enhancement framework in this paper.


In constructing the knowledge system for LLM-based expert systems, this paper creates a structured JSON-format external knowledge base. JSON’s inherent facilitation of data extraction and processing, relative to alternative formats, establishes a robust foundation for retrieving broader and more precise knowledge during search operations\cite{dong2025leveragingllmassistedqueryunderstanding}. The knowledge base integrates power domain expert knowledge and tool documentation, including system optimization models, devices, objective functions, constraints, and support for dynamic updates. Additionally, it includes detailed descriptions of the core classes, functions, and constraints in the cvxpy library \cite{diamond2016cvxpy}. Furthermore, this paper designs a multi-level prompt engineering strategy, which first clarifies the roles of various domain experts, then uses Chain-of-Thought techniques to guide experts in step-by-step reasoning, and ensures that the LLM structures content according to predefined rules.

\subsubsection{Prosumer-Centric LLM Expert Strategies}

This paper presents an LLM-based expert execution workflow to support personalized operational strategy for each prosumer. The system includes four complementary LLM agent experts and a module for distribution networks security verification via DSO, All optimization models are formulated in cvxpy and solved using mathematical optimization solvers such as Gurobi \cite{gurobi}. The overall process is shown in Fig. \ref{fig:LLM}, as described below:

\begin{figure*}[!t]
    \includegraphics[width=1\linewidth]{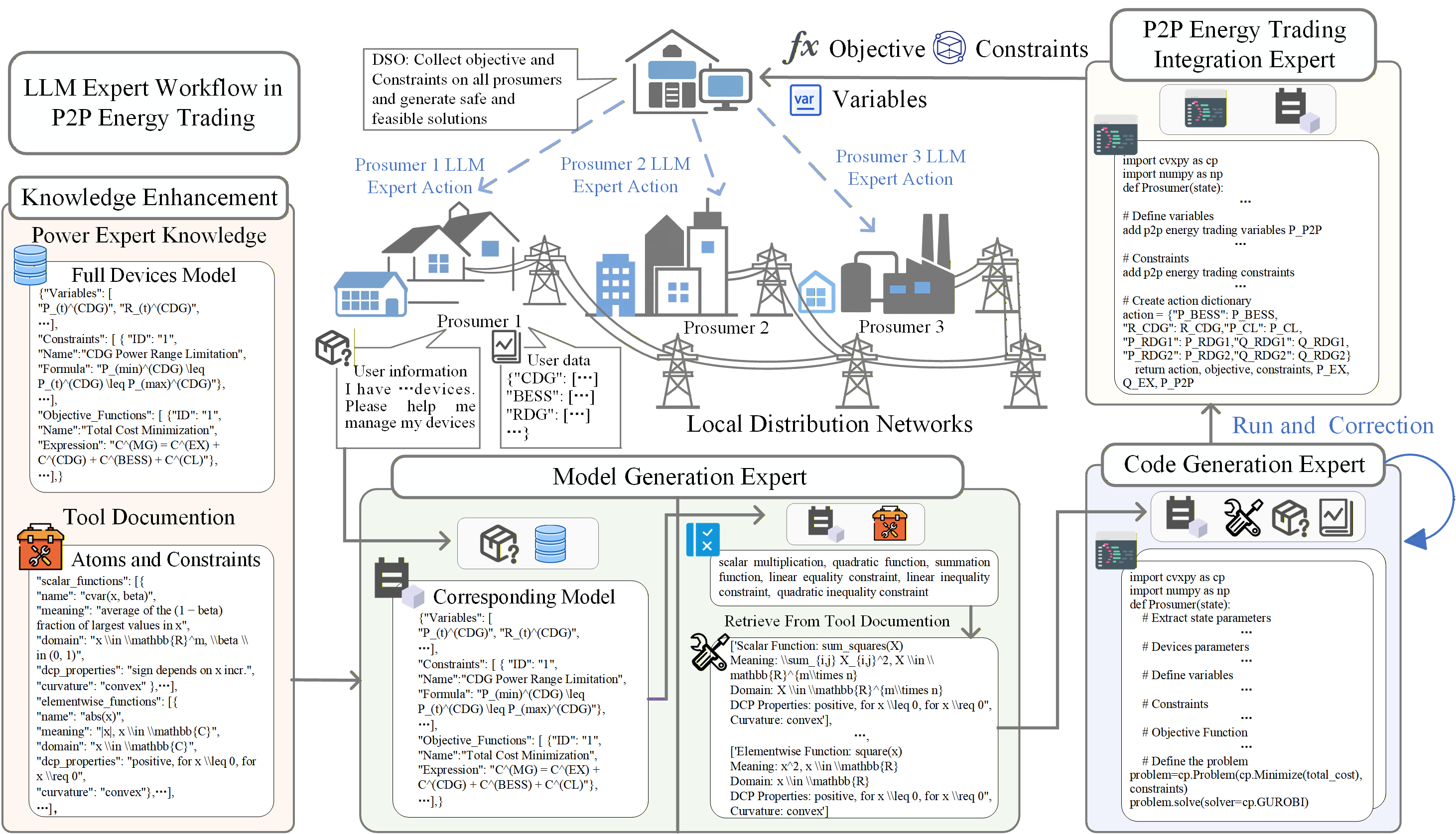}
    \caption{Our proposed LLM expert workflow in P2P energy trading}
    \label{fig:LLM}
\end{figure*}







\begin{itemize}

\item \textbf{Model Generation Expert:} This module LLM extracts key devices and optimization requirements from the input of the prosumer's natural language, generates the corresponding model knowledge, and predicts relevant cvxpy Atoms and Constraints based on retrieval results and tool documentation.

\item \textbf{Code Generation Expert:} This module LLM constructs the optimization model in the cvxpy framework using knowledge of the power domain, the cvxpy programming syntax, and the device parameters and state data of the prosumer. It outputs DCP-rule-compliant and feasible modeling code.

\item \textbf{Iterative Correction Expert:} This module LLM runs the model in a sandbox environment using cvxpy together with commercial solvers, detecting and correcting syntax errors and runtime issues, ensuring that the model is executable and complete.

\item \textbf{Energy Trading Integration Expert:} After prosumer modeling and validation, this module LLM integrates the P2P energy trading variables into the optimization model, adding the necessary objective functions and constraints, and outputs personalized objective function, constraint list, and node injection of active and reactive power.

\item \textbf{Distribution Networks Security Verification:} {After all prosumer objectives and constraints are submitted, the DSO adds the objective functions and constraint sets of all prosumers. It then incorporates the active and reactive power node injections into the branch flow model}\cite{WOS:000322989900051} {and invokes a commercial optimization solver to verify the global power flow correction across the distribution network, thereby generating optimized operating strategies for the prosumers.}

\end{itemize}

\subsection{Multi-Agent Imitation Learning Algorithm}


For the prosumer collaborative learning problem, this paper proposes an expert strategy-constrained MARL Algorithm based on the CTDE framework. The Algorithm constructs a joint state-action value function (Q-function) and state value function (V-function) through a centralized evaluation network, integrating the global environmental state and the behavior policies of all agents during the training phase, thereby guiding the differential optimization of individual agent policies. During execution, a decentralized approach is adopted, where each agent independently makes decisions based on local observations through independently trained networks, balancing cooperative benefits and decision-making efficiency.

\subsubsection{Preparation}

The expert strategy constraints are embedded within a multi-agent actor-critic framework to solve the constrained optimization problem. Based on the Lagrangian dual theory, this problem is first transformed into its Lagrangian dual form. For each prosumer agent $i$, the multi-agent formulation can be expressed as:
\begin{equation}
    \begin{aligned}
  & \underset{\pi}{\text{min}} \ \mathbb{E}_{{s}_{t}\sim \mathcal{B},{a}_{t} \sim \pi_{\phi_{i}} ( \cdot |{{o}_{i,t}})} \Big[-\underset{z \in {1,2}}{\text{min}}Q_{\theta_{z}}(s_{t},{a}_{t}) \Big] \\ 
 & \text{s}\text{.t}\text{.} \ \begin{matrix}
   {}  \\
\end{matrix} {\hat{W}}_{2}\left( \pi_{\phi_{i}} ( \cdot |{{o}_{i,t}}),{{\pi }^{LLM}}( \cdot |{{o}_{i,t}}) \right) \le \epsilon \ ,  \\ 
\end{aligned}
\label{basic}
\end{equation}
where $\mathcal{B}$ is the experience replay buffer; $\phi$ is the policy network parameters and the output sampling from the Gaussian distribution $a_{i,t} \sim \mathcal{N}(\mu_{\phi_{i}},\sigma_{\phi_{i}}^{2})$; $\epsilon$ is the policy deviation; $\theta$ is the Q-function network parameters and z = 1,2; $\psi$ is the V-function network parameters. The policy network outputs a distribution over actions, which inherently facilitates exploration during training by enabling diverse action sampling under identical states.

Since the LLM-based expert strategies can only generate the mean parameters of the policy distribution, without estimating the standard deviation, the output is modeled as a Dirac delta function representing a degenerate distribution. To measure the similarity during policy iterations, the Wasserstein-2 metric, known for its distributional robustness, is used. In the case of a one-dimensional Gaussian distribution, ${\hat{W}}_{2}^{2}$ has a closed-form analytical solution\cite{COTFNT}:
\begin{equation}
    {\hat{W}}_{2}^{2}( \pi_{\phi_{i}} (\cdot |{{o}_{i,t}}),{{\pi }^{LLM}}(\cdot |{{o}_{i,t}})=\int_{0}^{1}\left|{F}^{-1}(q)-G^{-1}(q)\right|^{2}d q,
\end{equation}
\begin{equation}
    \begin{split}
        \int_{0}^{1}&\left(\mu_{\phi_{i}}+\sigma_{\phi_{i}}\Phi^{-1}(q)-a_{i,t}^{LLM}\right)^{2}d q 
        = \int_{0}^{1}\Big[(\mu_{\phi_{i}}-a_{i,t}^{LLM})^{2} 
        \\ 
        & + 2(\mu_{\phi_{i}}-a_{i,t}^{LLM})\sigma_{\phi_{i}}\Phi^{-1}(q) 
        + \sigma_{\phi_{i}}^{2}\left(\Phi^{-1}(q)\right)^{2}\Big]d q.
    \end{split}
\end{equation}

The resulting Wasserstein-2 metric between the expert strategy and the policy network is:
\begin{equation}
   {\hat{W}}_{2}( \pi_{\phi_{i}} (\cdot |{{o}_{i,t}}),{{\pi }^{LLM}}(\cdot |{{o}_{i,t}})=\sqrt{(\mu_{\phi_{i}}-a_{i,t}^{LLM})^{2}+\sigma_{\phi_{i}}^{2}}.
   \label{wass}
\end{equation}

\subsubsection{Multi-Head Differential Attention}
To enhance the modeling capability of the centralized critic in capturing inter-agent interactions, a multi-head differential attention mechanism is integrated into the critic network. This design aims to improve the accuracy of global value estimation. For each agent $i$, the input is first transformed into an embedding $e_{i}$ and then $E$ represents the traversal of all other agents embedded vectors. Matrix $E$ is then split into $h$ attention heads and linearly projected into the query, key, and value spaces for each head, as follows:
\begin{equation}
[Q^{h}_{1}, Q^{h}_{2}] = e_{i}W^Q,\ [K^{h}_{1}, K^{h}_{2}] = EW^K,\ V^{h} = EW^V,
\end{equation}
where $Q^{h}_{1}, Q^{h}_{2}, K^{h}_{1}, K^{h}_{2}, V^h \in \mathbb{R}^{d_{\text{model}}}$. The differential attention mechanism operates by subtracting two softmax-based attention maps, aiming to eliminate redundant information among agents and emphasize critical dependencies. The attention output for each head is computed as:
\begin{equation}
\begin{aligned}
  \text{head}^{h}= \bigg[\text{softmax(}\frac{{{Q}^{h}_{1}}K_{1}^{h\top }}{\sqrt{{{d}_{k}}}}\text{)}-\xi^{h}\text{softmax(}\frac{{{Q}^{h}_{2}}K_{2}^{h\top }}{\sqrt{{{d}_{k}}}}\text{)}\bigg ]{{V}^{h}}, 
\end{aligned}
\end{equation}
\begin{equation}
    X = \text{Concat(head}^{1},\cdots \cdots, \text{head}^{h}\text{)}W^{O},
\end{equation}
where $W^{O}$ is a project matrix; scaling factor $\xi^{h}$ is a learnable scalar and dynamically computed as:
\begin{equation}
\xi^{h} = \exp(\xi_{\mathrm{q_1}} \cdot \xi_{\mathrm{k_1}}) - \exp(\xi_{\mathrm{q_2}} \cdot \xi_{\mathrm{k_2}}) + \xi_{\mathrm{init}},
\end{equation}
where $\xi_{\mathrm{q_1}}, \xi_{\mathrm{q_2}}, \xi_{\mathrm{k_1}}, \xi_{\mathrm{k_2}}$ are trainable vectors that vary with the head index.

\begin{figure}[!t]
    \centering
    \includegraphics[width=0.8\linewidth]{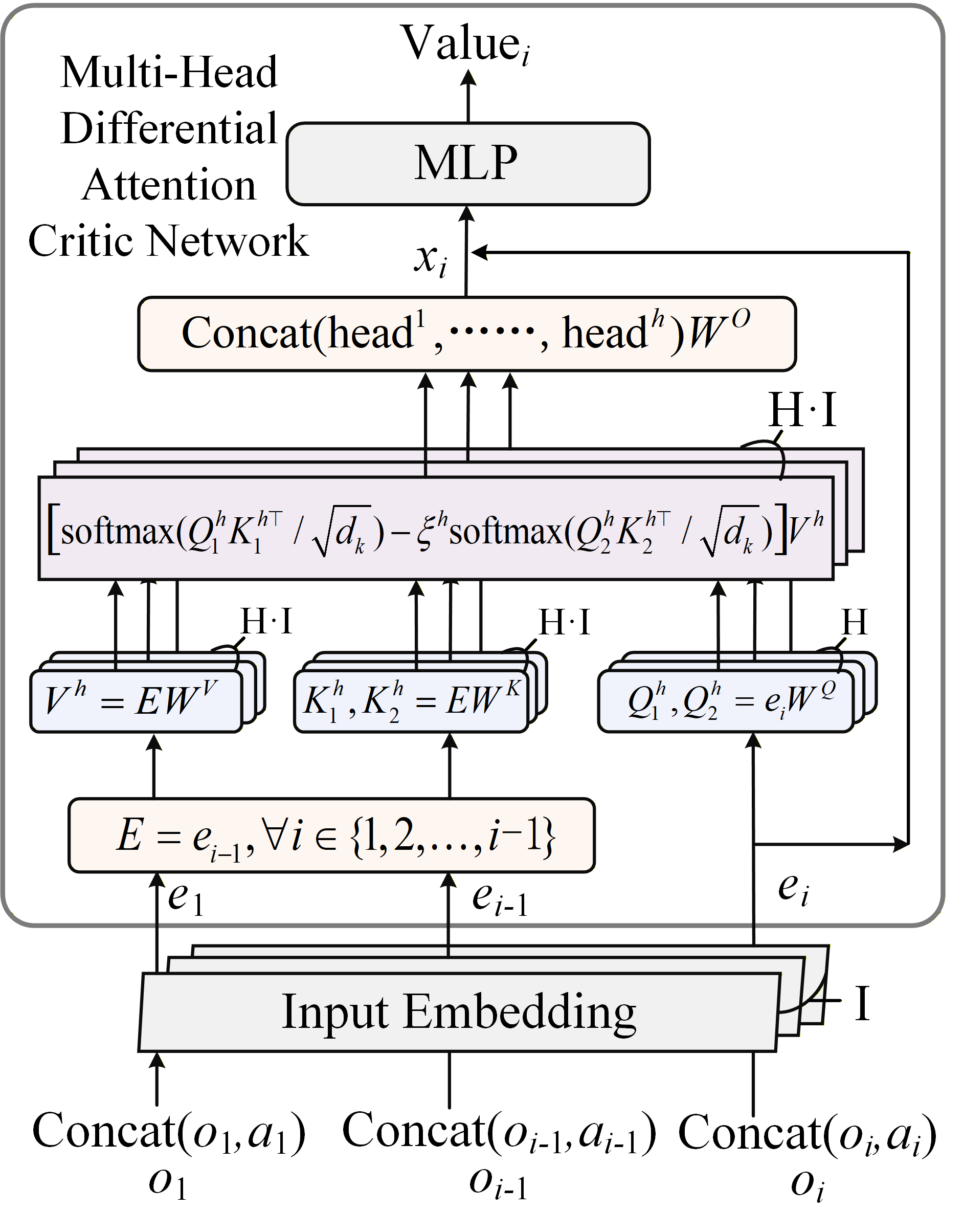}
    \caption{Multi-head differential attention critic network}
    \label{fig:attention}
\end{figure}

\hl{To enhance the representational capacity of critic networks while preserving ego-specific information, we employ a feature fusion mechanism that concatenates the intermediate output} $x_i$ \hl{with the original agent embedding} $e_i$. \hl{This ensures that agent-specific features remain directly accessible to subsequent layers, mitigating potential information loss during intermediate transformations. The fused representation is processed by an MLP of fixed dimensionality, guaranteeing that its size remains invariant to agent population growth. This fusion facilitates flexible feature interactions and refines the representation prior to generating the critic's final value estimate. The overall architecture is illustrated in Fig. }\ref{fig:attention}.

{The differential attention mechanism is motivated by the observation that, in real-time P2P energy trading, only a subset of prosumers significantly influences an agent’s decision. For example, when an industrial prosumer exhibits high energy demand, cooperating with nearby prosumers possessing renewable surpluses can significantly enhance the social welfare. In contrast, during low-demand periods, such trading interactions are less influential to the cooperative outcome. Conventional attention mechanisms may assign non-negligible weights to local irrelevant prosumers, thereby diluting focus on critical trading relationships. To address this, the proposed mechanism computes two attention maps, one capturing general relevance and the other suppressing redundant interactions, and subtracts them. This operation sharpens the critic network's focus on discriminative inter-agent dependencies, enabling more accurate value estimation.}

\subsubsection{Learning the Critics}


In scenarios where agents share a global reward, a major challenge is to reduce the variance of policy gradient estimates in interactive multi-agent environments. To address this issue, a double Q-function network and a target V-function network are employed. The minimum selection operation in the double Q-function network effectively mitigates overestimation bias. The incorporation of a V-function network contributes to a significant reduction in estimation variance \cite{Lyu_Baisero_Xiao_Amato_2022}. All agents share critic networks with shared parameters; this design significantly reduces the number of trainable parameters and improves training stability and scalability in large-scale multi-agent systems.

For V-function network $V_{\psi}$, the loss function is calculated to approximate the given the current state and Q-function values:
\begin{equation}
    \mathcal{L}_{V}(\psi)=\mathbb{E}_{s_{t}\sim \mathcal{B},{a}_{t}\sim\pi_{\phi_{i}} ( \cdot |{{o}_{i,t}})}\Bigl[(V_{\psi}({s}_{t})-\underset{z \in {1,2}}{\text{min}}Q_{\theta_{z}}(s_{t},{a}_{t}))^{2}\Bigr].
\end{equation}

For the Q-function network $Q_{\theta}$, the training target $y_t$ is computed using the a delay updated target network $V_{{ \overline{\psi}}}$.  The loss function is defined as follows:

\begin{equation}
    \begin{split}
        \mathcal{L}_{Q}(\theta) =& \mathbb{E}_{(s_{t},,r_{t},s_{t+1})\sim\mathcal{B},a_t \sim \pi_{\phi_{i}}( \cdot |{{o}_{i,t}})} \biggl[ 
        \frac{1}{2} \left( Q_{\theta}(s_{t}, a_{t}) \right.
        \left. - y_{t} \right)^2 \biggr]\\
        &y_{t}=r_{t}+\gamma V_{\bar{\psi}}(s_{t+1}) \ , 
    \end{split}
\end{equation}
where the target V-function network ${{ \overline{\psi}}}$ is updated via  Polyak averaging rather than direct copying to enhance stability:
\begin{equation}
\bar{\psi} \leftarrow \tau \psi + (1-\tau)\bar{\psi} \ ,
\end{equation}
where $\tau \ll 1$ is the smoothing coefficient. 

\subsubsection{Learning the Actors}
Following the critic network update, the policy constraint in (\ref{basic}) can be expressed as a Lagrangian dual problem \cite{9694460}, where Lagrangian multipliers $\lambda_{i}>0$. The formulation becomes:
\begin{equation}
    \begin{split}
        \max_{\lambda} \min_{\pi} \ 
        &\mathbb{E}_{s_t \sim \mathcal{B}, a_t \sim \pi_{\phi_{i}} ( \cdot |{{o}_{i,t}})}  \biggl[ 
         -\min_{z \in \{1,2\}} Q_{\theta_{z}}(s_t, a_t)+ \\
        & \lambda_{i}\left( {\hat{W}}_2 \left( \pi_{\phi_{i}} (\cdot | o_{i,t}), \pi^{LLM} (\cdot | o_{i,t}) \right) - \epsilon \right) \biggr].
    \end{split}
\end{equation}

Policy improvement serves to optimize and update the MARL policy. $\phi_i$ updated by minimizing the following loss function:
\begin{equation}
    \begin{split}
        \mathcal{L}_{\pi}(\phi_{i}) = \ 
        &\mathbb{E}_{s_t \sim \mathcal{B}, {a_t \sim \pi_{\phi_{i}} ( \cdot |{{o}_{i,t}})}}\biggl[
         -\min_{z \in \{1,2\}} Q_{\theta_{z}}(s_t, a_t)+ \\
        & \lambda_{i} \left( {\hat{W}}_2 \left( \pi_{\phi_{i}} (\cdot | o_{i,t}), \pi^{LLM} (\cdot | o_{i,t}) \right) - \epsilon \right) \biggr].
    \end{split}
\end{equation}

By updating $\lambda_{i}$, the degree of constraint violation can be mitigated. This is achieved by minimizing the following loss function:
\begin{equation}
    \mathcal{L}(\lambda_{i})=\mathbb{E}_{s_{t}\sim \mathcal{B}}\left[ 
-\lambda_{i} \big({\hat{W}}_{2}( \pi_{\phi_{i}} (\cdot |{{o}_{i,t}}),{{\pi }^{LLM}}(\cdot |{{o}_{i,t}}\big)-\epsilon)  \right].
\end{equation}

In the initial phase of training, a low policy deviation is employed to guide the agent’s learning; during the middle and later phases, a larger policy deviation is introduced to sustain the agent’s exploration. This framework ultimately enables efficient and stable policy optimization.

\subsubsection{Prioritized Experience Replay}
A sample loss-based priority evaluation mechanism is introduced. Samples with higher losses are considered more valuable for the learning process and are assigned higher priority, increasing their sampling probability.


Two experience replay buffers are defined: the normal operation experience replay buffer stores experiences collected during time steps where the agent’s actions do not cause voltage violations, reflecting typical operating conditions; the constraint violation experience replay buffer stores experiences from extreme renewable energy time steps in which the agent’s actions lead to grid voltage limit violations. During training, experiences are sampled from both buffers separately with a loss-based priority evaluation mechanism, drawn in a ratio of $k:(1-k)$, to form training batches.

\section{NUMERICAL STUDY}
\begin{figure}
    \centering
    \includegraphics[width=0.8\linewidth]{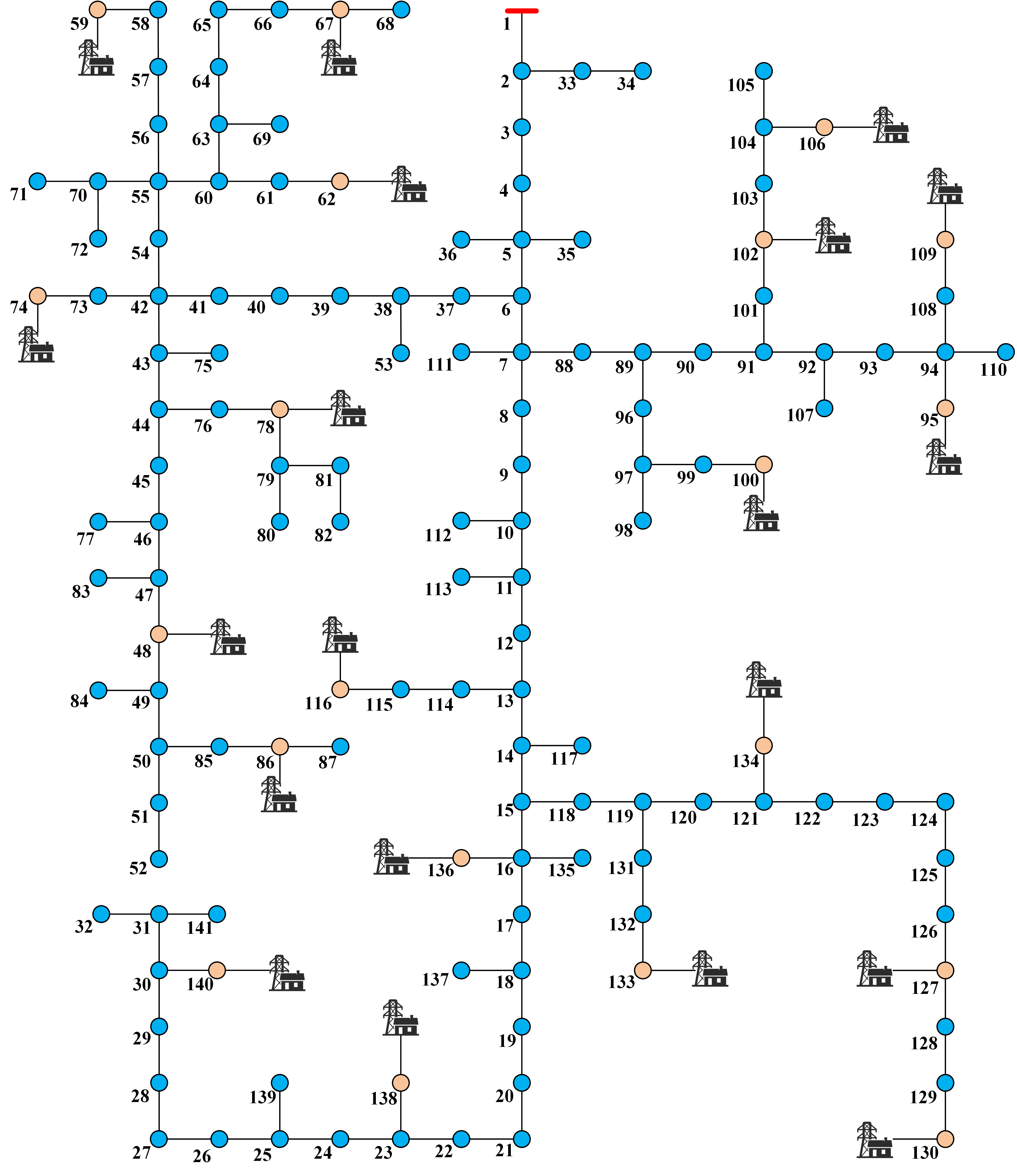}
    \caption{IEEE141‑bus distribution networks with twenty prosumers}
    \label{fig:ieee141}
\end{figure}

To validate the effectiveness of the proposed framework, a numerical study is carried out on a modified IEEE 141‑bus distribution network, the voltage level is 12.47kV. As shown in Fig. \ref{fig:ieee141}, 20 prosumers are selected to participate in local P2P energy trading, and the characteristics of five prosumer types are summarized in TABLE \ref{tab:heterogeneous-prosumers}.  The renewable generation outputs and load demand curves of the prosumers are extracted from a real‑world dataset published by the Belgian Transmission System Operator Elia\cite{EliaGridData}.

\begin{table}[htbp]
\centering
\adjustbox{max width=\linewidth}{ 
\begin{tabular}{@{}cl*{5}{c}@{}}
\toprule
\multirow{2}{*}{Node} 
& \multicolumn{5}{c}{Devices Portfolio} 
& \multirow{2}{*}{\begin{tabular}[c]{@{}c@{}}Prosumer\\ Scenario\end{tabular}}\\
\cmidrule(lr){2-6}
& CDG & WT & PV & BESS & CL  \\
\midrule
48,78,102,127,  & \checkmark & -- & \checkmark & \checkmark & \checkmark & Commercial \\
59,109,130,140 & -- & \checkmark & \checkmark & \checkmark & \checkmark & Rural \\
67,95,133,136 & \checkmark & \checkmark & -- & -- & \checkmark & Industrial \\
62,86,106,138 & -- & -- & \checkmark & \checkmark & \checkmark & Residential \\
74,100,116,134 & \checkmark & \checkmark & \checkmark & \checkmark & \checkmark & Energy Hub \\
\bottomrule
\end{tabular}%
}
\caption{Personalized configurations for 20 prosumers}
\label{tab:heterogeneous-prosumers}
\end{table}

\subsection{Comparison Baselines}
To evaluate the performance of the proposed algorithm, we compare it against the following baselines while also introducing:

\begin{enumerate}[]
\item MADDPG\cite{bettini2024benchmarlbenchmarkingmultiagentreinforcement}: A CTDE-based multi-agent extension of DDPG employing a centralized critic and decentralized actors.

\item MAAC \cite{iqbal2019actorattentioncriticmultiagentreinforcementlearning}: A CTDE framework augmented with a soft attention mechanism that adaptively weights and filters inter-agent information.

\item MATD3+BC\cite{fujimoto2021minimalistapproachofflinereinforcement}: Enhances MATD3 by incorporating a behavior-cloning loss to align each agent’s policy with LLM-generated expert actions.

\item MAGAIL\cite{song2018multiagentgenerativeadversarialimitation}: \hl{A multi-agent extension of GAIL, in which an adversarial discriminator is trained to distinguish expert trajectories generated by LLM from the agents’ joint state–action trajectories. The output of the discriminator is subsequently transformed into a shaped imitation reward, which replaces the environment reward.}

\item \hl{Our Proposed: Introduces a Lagrange multiplier during training to progressively constrain agent behaviors toward the LLM expert’s strategies.}

\item Our Proposed-MH: Extends the Our Proposed algorithm by integrating a differential multi-head attention mechanism into the critic to improve global value estimation.
\end{enumerate}

\subsection{Implementation Details}

{In terms of neural network architecture design, aside from the variants incorporating attention mechanisms, all baseline algorithms share a unified network architecture under identical hyperparameter settings. The detailed hyperparameter settings are provided in TABLE }\ref{tab:hyperparameters}. Training was conducted for 5000 episodes with the same random seed initialization and update frequency. All experiments were implemented in Python 3.11.10 under the PyTorch 2.7, with parameters updated via the Adam optimizer. Computations were performed on a platform equipped with an NVIDIA RTX 5070Ti GPU, an AMD Ryzen ThreadRipper 3970X CPU, and 64GB of RAM.

\begin{table}[ht]
	\centering
	\caption{Hyperparameters used in the MARL}
	\label{tab:hyperparameters}
	\begin{tabular}{@{}ccc@{}}
		\toprule
		Symbol & Meaning & Value \\ \midrule
		$lr$ &Learning rate &1e-4\\
		$N_{episode}$ &Maximum episode & 5000 \\
		$\gamma$ & Discount factor & 0.99 \\
		$N_{B}$ & Replay buffer size & 1e5 \\
		$\mathcal{B}$ & Training batch size & 128 \\
		$\lambda$ & Initial Lagrange Multiplier  & 0.02 \\
		$\epsilon$ & Policy deviation & 0.1 \\
		$k$ & Proportion of normal replay buffer & 0.8\\
		\bottomrule
	\end{tabular}
\end{table}


Using year-round data, we designate the first day of each month as the validation set, the last week of each month as the test set, and the remaining days for training. A dynamic validation strategy is employed during training. After each training step, the current policy is evaluated using the validation set. The average cumulative discounted reward across all validation sets is recorded as a performance metric. To ensure statistical robustness, each algorithm is executed independently five times, and both the mean reward and the standard deviation of the resulting rewards are recorded.

In the present study, the large parameter LLM workflow is implemented through online API interfaces for experimental validation. Nevertheless, the proposed architecture is fully modular, allowing seamless substitution with locally hosted or SFT LLM instances to ensure data privacy and reduce latency. \hl{The LLM workflow is activated after every 200 episodes during the MARL training phases or system reconfiguration stages (e.g., when new devices are added or topology changes occur). This approach effectively mitigates the decline in learning performance caused by any poor generation by the LLM in a given episode.} During every training step, each prosumer executes the LLM pre-generated optimization model code in parallel, while the DSO merely aggregates the objective functions and constraints from all prosumers and utilizes a commercial solver to generate secure operational strategies for each prosumer. \hl{To address potential runtime burdens, efficient memory management is implemented to ensure the computational tractability of the iterative training process.} In real-time operation, lightweight MARL agents autonomously perform decision-making without additional LLM guidance or global DSO validation. \hl{In particular, when the LLM expert workflow encounters a device that is not available in the knowledge base, the policy deviation is deliberately increased to a very large value. Under this setting, the agent in our proposed algorithm that lacks valid LLM expert demonstrations effectively degenerates into a reward-driven reinforcement learning process, as the influence of expert guidance is implicitly suppressed.}

\subsection{Performance Comparison}
\subsubsection{Analysis of the LLM Expert Strategy}
{In this evaluation, the human expert baseline refers to the optimal strategy obtained by solving a deterministic convex optimization model—formulated in accordance with the problem setting described in Section II—using the Gurobi. This model fully incorporates all physical constraints of the prosumers and minimizes the total operational cost under perfect forecasts of renewable generation and load demand. The resulting solver-based optimal policy serves as the ground-truth reference for assessing the accuracy of LLM-generated strategies.}

Since the actions produced by the LLM expert directly influence subsequent MARL training, we evaluate the LLMs on the prosumer task using the following four key metrics: 

\begin{itemize}
\item Pass Rate: The success rate of error-free, executable outputs, reflecting the LLM's ability to generate valid code.
\item Accuracy: For successful executions, accuracy quantifies the similarity to human expert actions based on cost deviation and action gap:
\begin{small}
\begin{align}
&\text{Deviation} = \frac{|C^{\text{LLM}} - C^{\text{Human}}|}{C^{\text{Human}}} \times 100\%,\\
&\text{Gap} = \frac{1}{T} \sum_{t=1}^{T} \left|\frac{a^{\text{LLM}}_{t} - a^{\text{Human}}_{t}}{a^{\text{Human}}_{t}}\right| \times 100\%,\\
&\text{Accuracy} = 100\% - \frac{\text{Gap} + \text{Deviation}}{2}.
\end{align}
\end{small}

\item Correction: The average number of code-fix iterations required before a successful execution, indicating generation efficiency.
\item Tokens: The average number of completion tokens per successful run, reflecting the computational cost of generating expert policies.
\end{itemize}

The workflow is implemented using LangGraph \cite{langchain_inc_langgraph_nodate}, where the number of code generation iterations per execution cycle is capped at a maximum of 5 iterations. The temperature parameter is set to 0.5 to balance randomness and determinism in the model output. All experiments utilize the latest publicly accessible LLMs via official API interfaces. For each model, 10 experimental trials are conducted per type of prosumer request, resulting in 50 total trials per model. The detailed results are as follows:

\begin{table}[ht] 
\caption{Performance comparison of different LLMs in workflow without prosumer preference}
\label{tab:comparison_llm}
\centering
\small 
\setlength{\tabcolsep}{8pt} 
\adjustbox{max width=\linewidth}{
\begin{tabular}{@{}lcccc@{}}
\toprule
LLM & Pass Rate(\%) & Accuracy (\%) & Correction & Tokens \\ 
\midrule
Chatgpt-4o            & 88  & 92.76  & 1.38  & 4727 \\
Claude-3.5-Sonnet & 94 & 99.41  & 0.95  & 5929 \\
Gemini-2.5-Flash  & 92 & 98.65  & 1.24  & 18856 \\
DeepSeek-V3       & $\bm{96}$  & 96.45  & $\bm{0.43}$  & $\bm{4039}$ \\
Qwen-2.5-Max       & 90  & $\bm{99.62}$  & 1.54  & 5302 \\
\cdashline{1-5}
Chatgpt-o3     & $\bm{100}$  & 98.52  & 0.28  & 17195 \\
Claude-4-Opus & $\bm{100}$ & $\bm{99.93}$  & $\bm{0.20}$  &  9454 \\
Gemini-2.5-pro  & $\bm{100}$ & 99.86  & 0.48  & $\bm{7558}$ \\
DeepSeek-R1     & $\bm{100}$  & 98.31  & 0.54  & 16687 \\
Qwen-3       & $\bm{100}$  & 99.84  & 0.33  & 8157 \\
\bottomrule
\end{tabular}
}
\end{table}

{TABLE} \ref{tab:comparison_llm} {provides a detailed comparison of the four metrics proposed in this study, where the LLMs above the dashed line deactivates advanced reasoning capabilities, while those below activate it. The latency of a large model depends on the network speed when API calls are used, but is effectively eliminated when the model is deployed locally. Specifically, LLMs without advanced reasoning capabilities are generally more cost-efficient; among these, Google Gemini-2.5-Flash shows the highest average cost at 2.43 CNY (Derived from the official billing statements of different LLMs after running the workflow once, and converted into CNY based on the prevailing exchange rate), mainly because it generates a larger number of output tokens per query, leading to higher overall billing despite its relatively low per-token price. In contrast, among reasoning-enabled models, Claude 4-Opus is the most expensive, with an average price of 6.27 CNY, primarily due to its higher per-token pricing structure reflecting the premium positioning of its advanced reasoning capabilities. Open-source LLMs such as DeepSeek and Qwen can also completely eliminate API-related expenses when deployed locally.}

The proposed multi-agent framework demonstrates superior model compatibility, enabling seamless integration with diverse mainstream LLMs rather than being constrained to the performance of a single model. A key finding is that the workflow exhibits exceptional performance in LLMs with advanced reasoning capabilities, achieving a pass rate of 100$\%$ in the evaluated tasks; this underscores that while prompt engineering merely mimics the structure of reasoning, reinforcement learning effectively internalizes the genuine instinct for self-correction. Notably, Claude-4-Opus attains a level of proficiency that in these scenarios fully substitutes for human experts. The complete prompt for our proposed LLM workflow is publicly accessible in the supplementary materials \cite{jzk0806_suppfile_2025}.

\begin{figure}[ht]
\centering
\includegraphics[width=\linewidth]{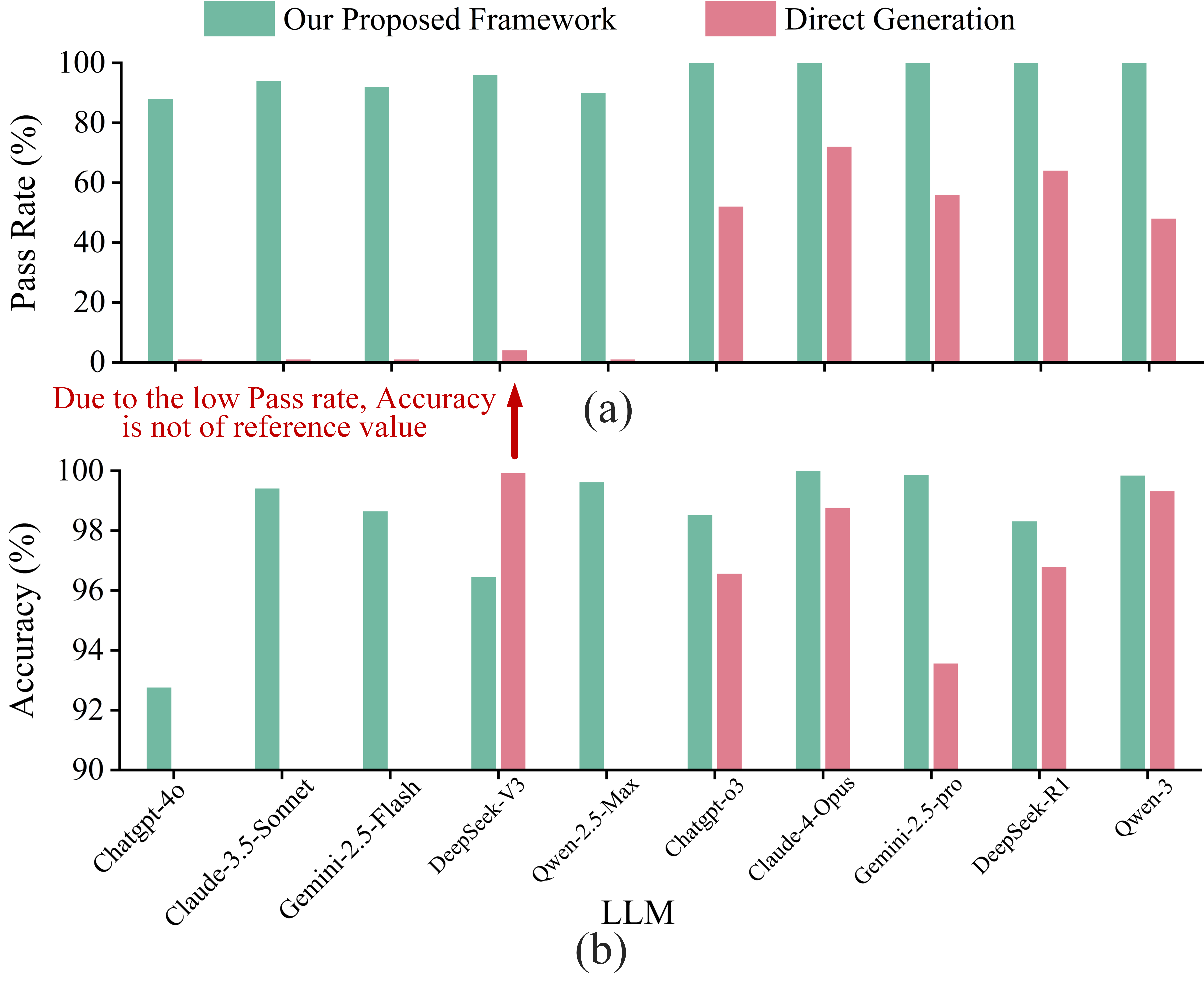}
\caption{Comparative experiments on the LLM framework without prosumer preference}
\label{fig:direct}
\end{figure}

As demonstrated in Fig. \ref{fig:direct}, the results validate the effectiveness of the proposed LLM-expert workflow. Comparative experiments were conducted where LLMs could not access Atomic Function retrieval content. In this scenario, LLMs were required to generate code directly based on input models, data, and user information, and then perform power flow verification without any iterative code correction. 
Cross comparisons of two core metrics—Pass Rate and Accuracy—reveal that models without the integrated framework exhibit severe performance degradation. For deactivated advanced reasoning  LLMs, pass rates approach zero due to primary failure modes such as "Model Infeasible or Unbounded" and "Numerical trouble encountered," indicating significant gaps from human-expert-level performance.

\begin{figure*}[!t]
    \centering
	\includegraphics[width=0.9\linewidth]{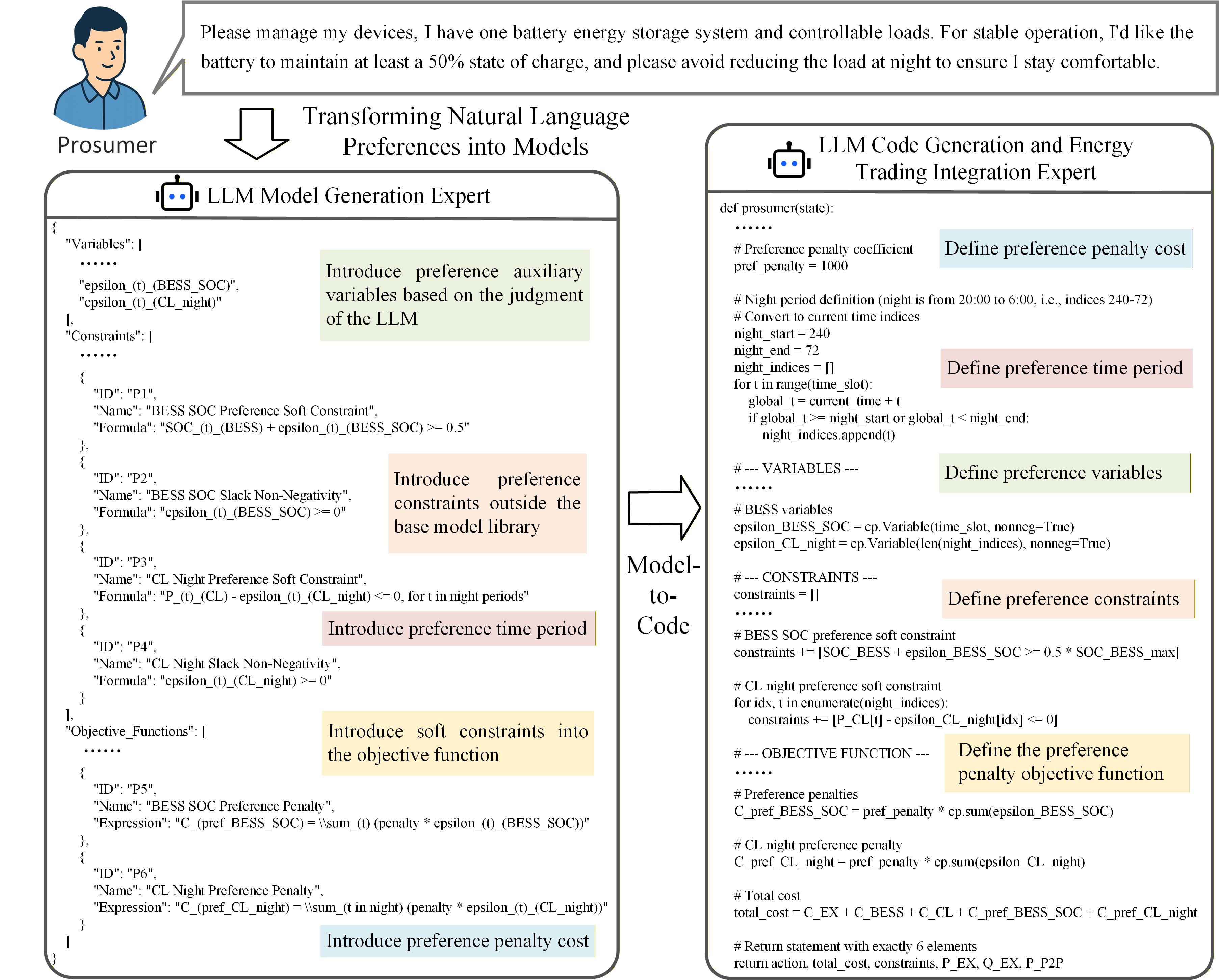}
	\caption{Our proposed LLM expert workflows for processing personalized prosumer preferences}
	\label{fig:LLM_preference}
\end{figure*}

\hl{To further investigate the performance of personalized prosumer preferences within the LLM expert workflow, we consider the potential conflicts between prosumer preferences and distribution network voltage constraints. Accordingly, the prompt instructs the LLM to formulate prosumer preferences as soft constraints in the objective function. This design ensures that distribution network security is always prioritized when conflicts arise, while prosumer preferences are satisfied to the greatest extent possible. The results are illustrated in Fig.} \ref{fig:LLM_preference}.

\begin{table}[ht]
	\caption{Performance comparison of different LLMs in workflow with prosumer preference}
	\label{tab:comparison_llm_pref_soft}
	\centering
	\small
	\setlength{\tabcolsep}{8pt}
	\adjustbox{max width=\linewidth}{
		\begin{tabular}{@{}lcccc@{}}
			\toprule
			LLM & Pass Rate(\%) & Accuracy(\%) & Correction & Tokens \\ 
			\midrule
			Chatgpt-4o            & 86 & 92.27 & 1.56 & 4904 \\
			Claude-3.5-Sonnet     & 92 & 99.14 & 1.01 & 6223 \\
			Gemini-2.5-Flash      & 90 & 98.34 & 1.47 & 19341 \\
			DeepSeek-V3           & $\bm{96}$ & 96.01 & $\bm{0.54}$ & $\bm{4362}$ \\
			Qwen-2.5-Max          & 88 & $\bm{99.39}$ & 1.75 & 5638 \\
			\cdashline{1-5}
			Chatgpt-o3            & $\bm{100}$ & 98.35 & 0.38 & 17773 \\
			Claude-4-Opus         & $\bm{100}$ & $\bm{99.81}$ & $\bm{0.28}$ & 13320 \\
			Gemini-2.5-pro        & $\bm{100}$ & 99.79 & 0.57 & $\bm{8365}$ \\
			DeepSeek-R1           & $\bm{100}$ & 98.01 & 0.65 & 17322 \\
			Qwen-3                & $\bm{100}$ & 99.66 & 0.43 & 9010 \\
			\bottomrule
		\end{tabular}
	}
\end{table}

\hl{TABLE} \ref{tab:comparison_llm_pref_soft} \hl{evaluates the performance of LLMs in handling prosumer preferences using 50 randomly generated, unambiguous profiles per prosumer type to ensure a fair comparison with human experts. Compared to TABLE} \ref{tab:comparison_llm}, \hl{the Pass Rate remains unchanged, while the Accuracy decreases slightly. This slight decline occurs because incorporating preferences increases the workflow complexity, necessitating more iterative refinements; consequently, both the Correction metric and Token count increase, although this does not compromise the LLMs' ability to successfully generate executable code. Further demonstrating the method's robustness, the results in Fig.} \ref{fig:direct_pref} \hl{highlight the advantages of the proposed workflow when handling the increased complexity of prosumer-defined optimization models. In contrast to Fig.} \ref{fig:direct}, \hl{omitting the proposed workflow in this setting leads to significant degradation: the lowest Pass Rate drops to only 30\%, while Accuracy falls to 85.43\%, proving the workflow's superior effectiveness in managing tasks with heightened complexity.}

\begin{figure}[ht]
	\centering
	\includegraphics[width=\linewidth]{Figure/llm2.png}
	\caption{Comparative experiments on the LLM framework with prosumer preference}
	\label{fig:direct_pref}
\end{figure}

\begin{table}[ht] 
	\caption{Performance Comparison of LLMs across Different Parameter in workflow with prosumer preference}
	\label{tab:comparison_llm_parameters}
	\centering
	\small
	\setlength{\tabcolsep}{8pt}
	\adjustbox{max width=\linewidth}{
		\begin{tabular}{@{}lcccc@{}}
			\toprule
			LLM & Pass Rate(\%) & Accuracy(\%) & Correction & Tokens \\ 
			\midrule
            \multicolumn{5}{c}{\textit{without preferences}} \\  
            Qwen-3-32B   & 82 & 89.34 & 2.21 & 6361 \\
			Qwen-3-14B   & 76 & 85.63 & 4.03 & 6908 \\
			Qwen-3-8B    & 44 & 76.85 & 1.27 & 6005 \\
			Qwen-3-4B    & 2  & 82.47 & 1.78 & 6187 \\
			\cdashline{1-5}
            \multicolumn{5}{c}{\textit{with preferences}} \\
			Qwen-3-32B   & 78 & 87.41 & 2.63 & 6398 \\
			Qwen-3-14B   & 64 & 85.08 & 4.17 & 7036 \\
			Qwen-3-8B    & 26 & 74.86 & 1.61 & 6121 \\
			Qwen-3-4B    & 0  & 78.12 & 1.74 & 6202 \\
			\bottomrule
		\end{tabular}
	}
\end{table}

\hl{To address API connection limitations, we evaluated locally deployed open-source Qwen-3 models with reasoning activated across various parameter scales in TABLE} \ref{tab:comparison_llm_parameters}, \hl{testing 50 tasks with user preferences and 50 without. Workflow effectiveness declines with model size; reducing the scale from 14B to 8B makes matching human-expert Pass Rate and Accuracy unattainable.}\hl{LangGraph workflow analysis reveals that models under 8B parameters consistently fail to transform input specifications. These models omit key user devices, preferences, and Model Generation Expert-derived constraints, while also failing to select appropriate RAG-retrieved Atomic Functions. Furthermore, these models have nearly lost the capability to rectify erroneous code. Consequently, in successful instances, the number of correction rounds and token consumption actually decrease due to the lack of iterative refinement. Thus, Qwen-3 models with at least 14B parameters are recommended for local deployment.}

{The proposed LLM workflow demonstrates strong scalability within the scope of its structured knowledge base. Specifically, the LLM-based expert workflow can effectively accommodate different prosumer types, provided that the corresponding device models are already included in the knowledge base (note that constructing a comprehensive model repository is beyond the scope of this work, although research institutes, large utility companies and tech giants are already making efforts to build model libraries}\cite{EPRI_OpenPowerAI_2025}).{ For each prosumer, the LLM dynamically generates personalized trading strategies based on its specific configuration. However, it is essential to recognize that when faced with requests involving device types or operational scenarios not covered by its knowledge base, the LLM may generate plausible but inaccurate or unsupported models or code, potentially compromising the accuracy and reliability of the resulting expert strategies.}

\subsubsection{Performance Analysis of MARL Algorithm}
Claude-4-Opus is used as the expert LLM for algorithm performance comparison in this section. It uses three indicators: daily average reward, average operational cost, and average voltage violation rate to compare the proposed algorithm with the baseline algorithms:
\begin{figure*}[!t]
    \centering
    \subfloat[The smoothed curves of the average rewards\label{figure1}]{
        \includegraphics[width=2.25in]{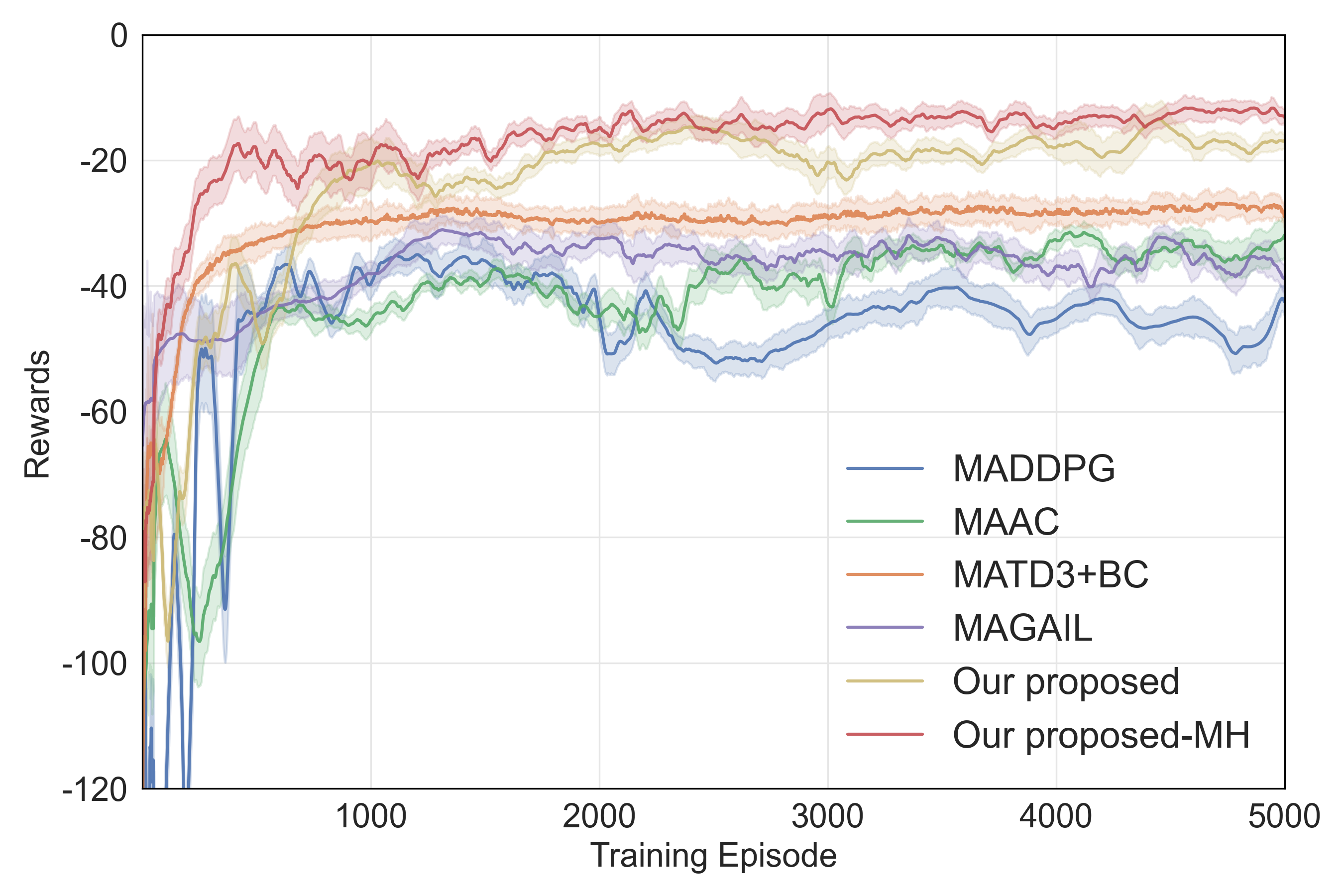}
    }
    \hfil
    \subfloat[The smoothed curves of the average operation cost\label{figure2}]{
        \includegraphics[width=2.25in]{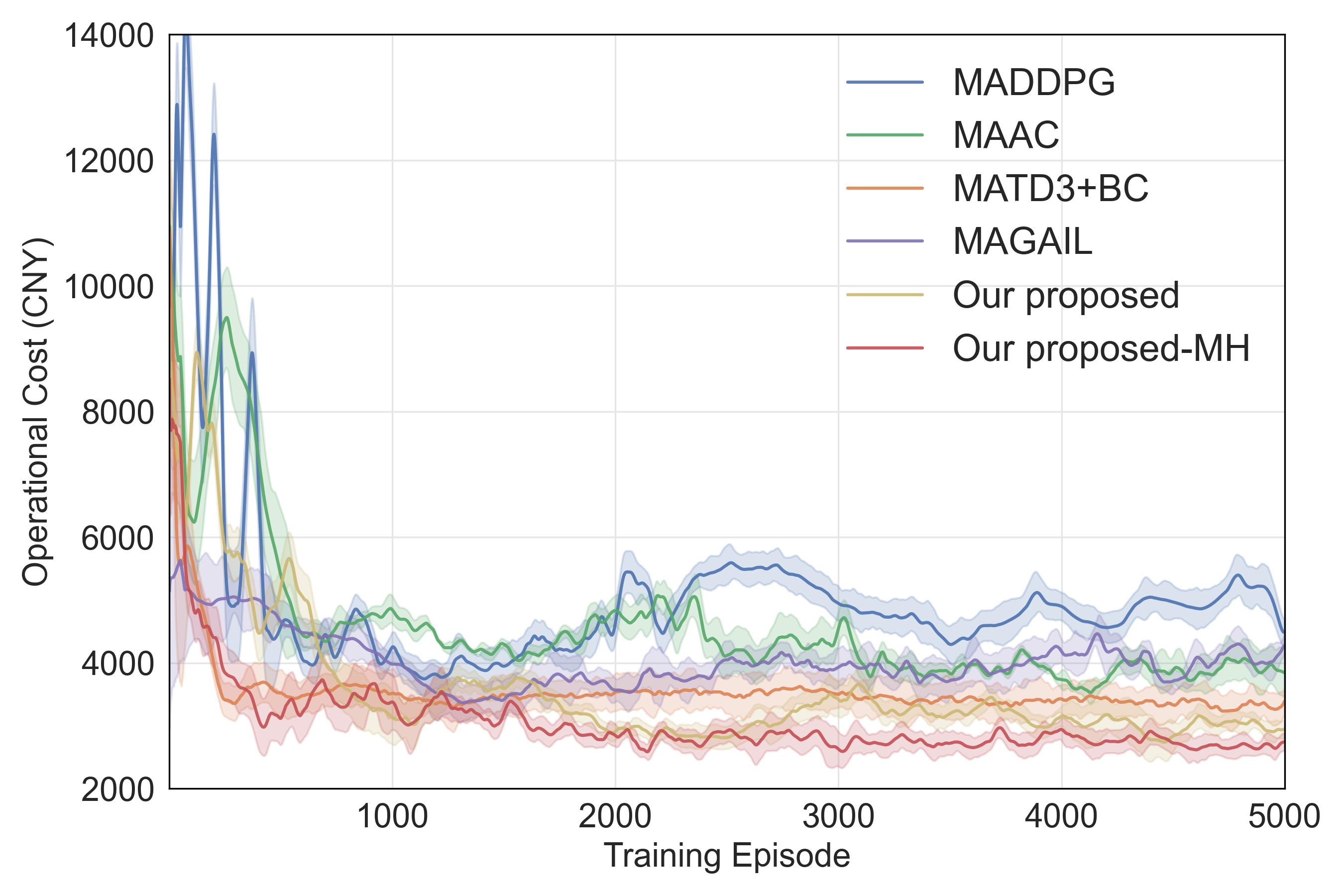}
    }
    \hfil
    \subfloat[The smoothed curves of the average voltage violation rate\label{figure3}]{
        \includegraphics[width=2.25in]{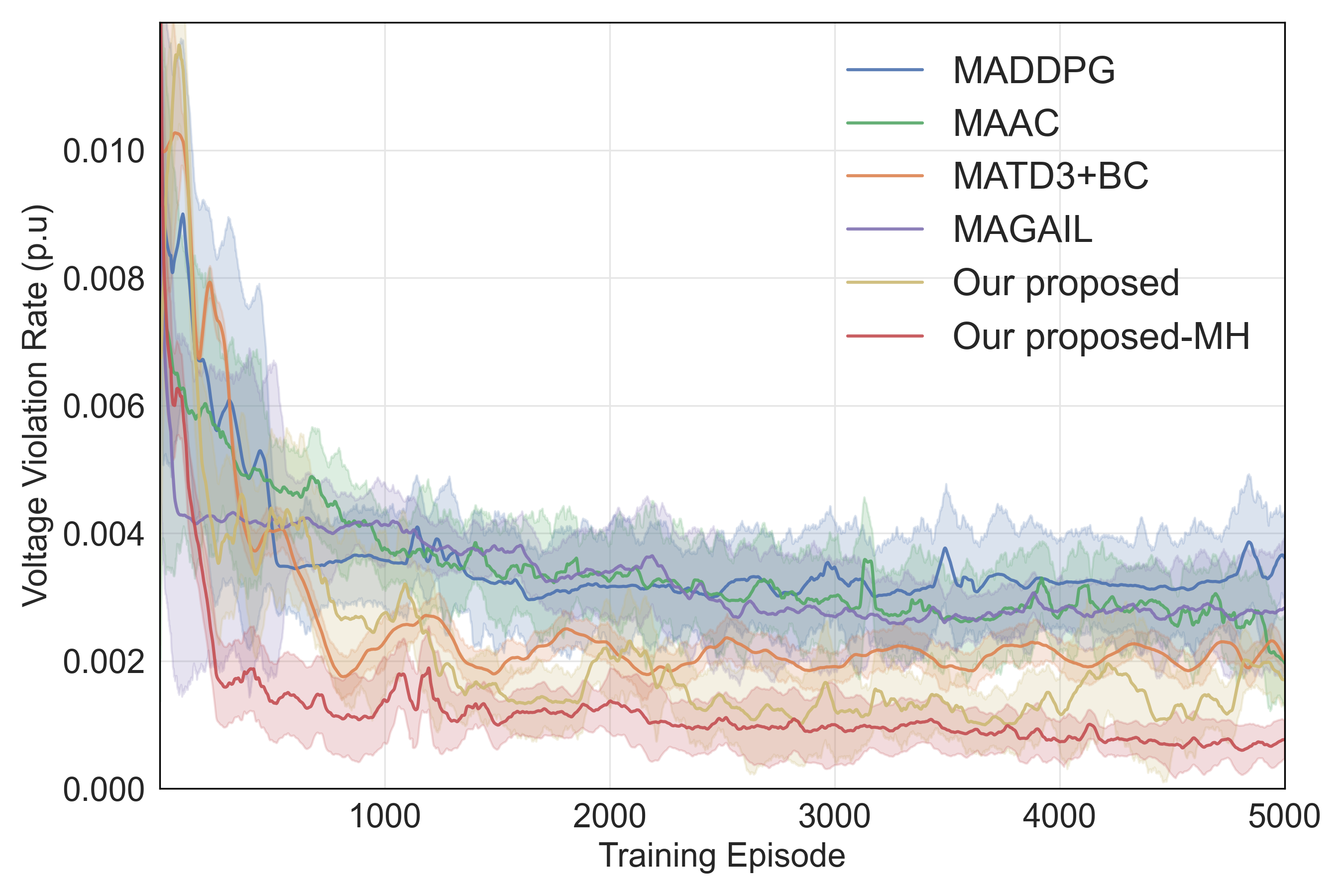}
    }
    \caption{Comparison chart of the performance of baseline algorithms}
    \label{baselines}
\end{figure*}

\hl{During the training of 20 prosumers, the proposed LLM expert workflow average replaced 3,039 lines of manually written human expert code.} The results in Fig. \ref{baselines} demonstrate the performance of the proposed algorithm during training on the validation set. As shown, 'Our proposed-MH' achieves faster convergence to a low operational cost with minimal fluctuations, demonstrating enhanced reward stability for guiding agents when LLM outputs expert actions. Furthermore, the curve of average voltage violation rate for 'Our proposed-MH' rapidly declines and maintains a low level, highlighting its strength in constraint satisfaction and ability to maintain secure grid operations.

\begin{table}[ht]
  \centering
  \caption{Comparison of different algorithms on test set}
  \label{tab:results}
  \adjustbox{max width=\linewidth}{%
    \begin{tabular}{@{}l 
                    S[table-format=4.2] 
                    S[table-format=3.2] 
                    S[table-format=1.2e-2] 
                    S[table-format=1.2e-2]@{}}
      \toprule
      \multirow{2}{*}{Algorithm}
        & \multicolumn{2}{c}{Operational Cost (CNY)}
        & \multicolumn{2}{c}{Voltage Violation Rate (p.u)} \\
      \cmidrule(lr){2-3} \cmidrule(lr){4-5}
        & {Mean} & {Std.\ Dev.} & {Mean} & {Std.\ Dev.} \\
      \midrule
     MADDPG          & 6586.45 & 340.76  & 3.66e-03 & 8.47e-04 \\
    MAAC            & 5940.32 & 220.49  & 2.02e-03 & 7.29e-04 \\
    MATD3+BC        & 5489.20 & 207.88  & 1.61e-03 & $\bm{2.82 \times 10^{-4}}$ \\
    MAGAIL          & 6380.96 & 164.23  & 2.82e-03 & 9.01e-04 \\
    OP    & 5146.27 & 146.07  & 1.39e-03 & 4.05e-04 \\
    OP-MH & $\bm{4840.61}$ & $\bm{135.31}$ & $\bm{1.06 \times 10^{-3}}$ & 3.03e-04  \\
      \bottomrule
      \multicolumn{5}{@{}l}{\textbf{Note:} OP stands for “Our Proposed.”}
    \end{tabular}%
  }
\end{table}
\sethlcolor{green}
{After completing model training, we applied the proposed algorithm to a test set to evaluate the practical applicability. As shown in TABLE}  \ref{tab:results}, {the proposed algorithm demonstrates significant advantages in both operational cost and voltage violation rate. Specifically, the average cost reached 4840.61 CNY, while the voltage violation rate was $1.06\times 10^{-3}$. Through comprehensive analysis of mean-standard deviation comparisons with baseline algorithms, the proposed approach achieves optimal results across all three evaluation metrics. It maintains the lowest mean values while exhibiting minimal standard deviations, indicating not only superior average performance but also robust stability.} 


\begin{figure}[ht]
	\centering
	\includegraphics[width=0.75\linewidth]{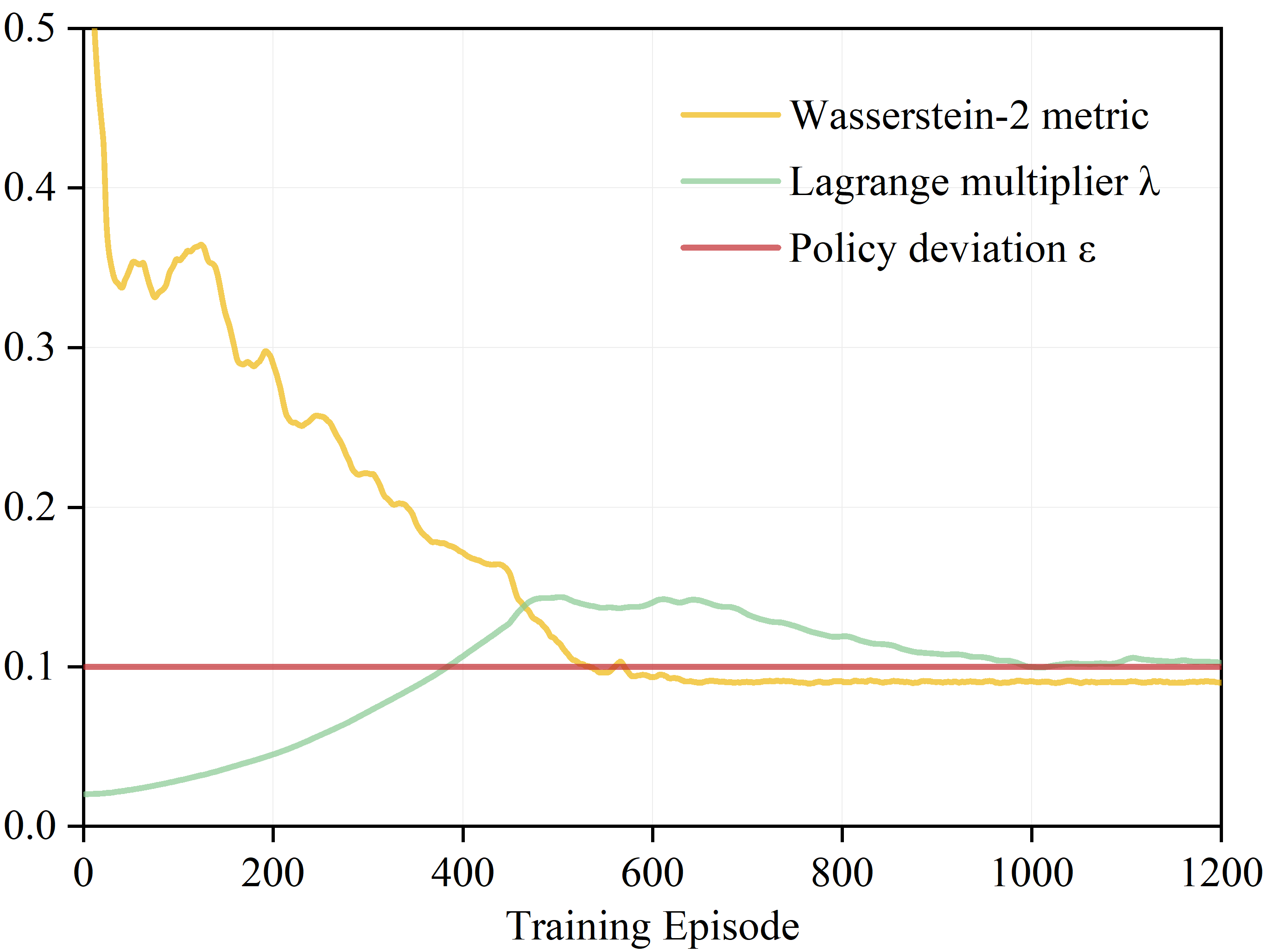}
	\caption{Average Lagrange multiplier and Wasserstein metric during training episode}
	\label{fig:lamda}
\end{figure}
\sethlcolor{white}
{During the training process, the evolution of the agent average Lagrange multiplier $\lambda$ and the average Wasserstein-2 metric provides key insight into the dynamic balance between expert imitation and autonomous policy optimization. As shown in Fig. }\ref{fig:lamda}, {$\lambda$ initially increases rapidly, reflecting the strong constraint effect imposed by the LLM expert strategy to guide the early-phase exploration of the MARL agents. With the gradual improvement of policy learning, $\lambda$ subsequently decreases and eventually stabilizes, indicating that the agents have sufficiently aligned with the expert behavior and require less external constraint. In contrast, the Wasserstein-2 metric between the agent policy and the expert policy exhibits a consistently decreasing trend, demonstrating continuous convergence of the learned policy toward the expert distribution. Once the Wasserstein-2 metric approaches the predefined threshold Policy deviation $\epsilon$, fluctuations remain bounded, signifying that the imitation constraint has reached equilibrium and the policy optimization process has entered a stable regime.}

\begin{figure}[ht]
	\centering
	\includegraphics[width=0.75\linewidth]{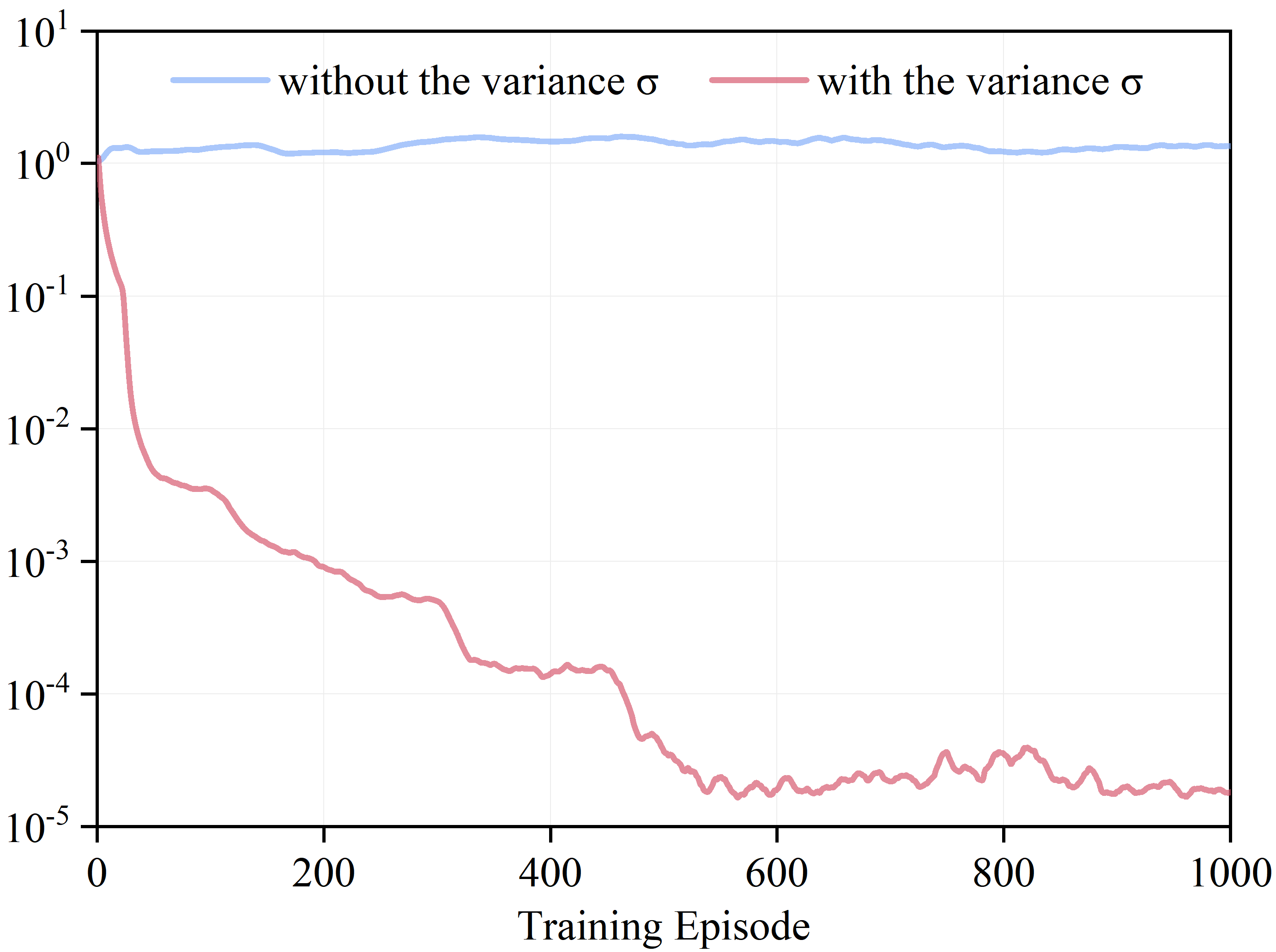}
	\caption{Sensitivity analysis of Wasserstein metric variance term}
	\label{fig:sigma}
\end{figure}
{To further investigate the sensitivity of the Wasserstein-based regularization, a comparative analysis was conducted between two formulations of the Wasserstein-2 metric in} (\ref{wass}): {one with the variance term $\sigma$ and one without it. As illustrated in Fig. }\ref{fig:sigma}, {when the $\sigma$ is incorporated, it converges smoothly to approximately $10^{-5}$ after around 500 episodes. This indicates that the stochastic exploration of the policy network is gradually reduced as the learned policy approaches the expert strategy, achieving a stable and near-deterministic policy behavior. In contrast, when the $\sigma$ is excluded, the implied policy variance remains large and fluctuates around 1.2 throughout training, suggesting incomplete convergence, the agent's actions are not stable enough. These results highlight that explicitly learning and regularizing the $\sigma$ is crucial for maintaining exploration in the early phase while ensuring eventual convergence to a stable policy distribution.}

\hl{To investigate the prosumer preference issue in P2P energy trading, heterogeneous preferences are assigned to 20 prosumers on the IEEE 141-bus topology shown in Fig. } \ref{fig:ieee141}. \hl{For each prosumer’s personalized preference, Our proposed method is guided by an LLM-based expert workflow.} 

\begin{figure}[ht]
	\centering
	\includegraphics[width=1.05\linewidth]{Figure/pref_comp.png}
	\caption{The operate results of our proposed algorithm on BESS and CL preference prosumer}
	\label{fig:pref_explore}
\end{figure}

\hl{The results in Fig. } \ref{fig:pref_explore} \hl{present a comparative analysis for prosumers exhibiting preferences toward BESS and CL operation under the proposed method and MAAC. Conventional MARL algorithms, represented by MAAC, are limited to conventional optimization behavior due to their strict reliance on the predefined reward function. In contrast, the proposed method further aligns the learned policy with expert strategies beyond purely reward-driven optimization, thereby enabling the consideration of prosumer preferences while simultaneously maintaining the maximization of social welfare.}

\hl{When the device type requested by a prosumer is not included in the knowledge base, or in local deployment settings, the use of LLMs with fewer than 8B parameters may lead to suboptimal expert actions. A comparative experiment is conducted between the proposed method and the existing baseline algorithm under a Qwen-3-8B model. After 1,000 training episodes, we explicitly schedule an increase in the policy deviation to allow the algorithm to perform autonomous exploration, while enforcing a lower bound policy variance to prevent premature policy determinism.}

\begin{figure}[ht]
	\centering
	\includegraphics[width=0.95\linewidth]{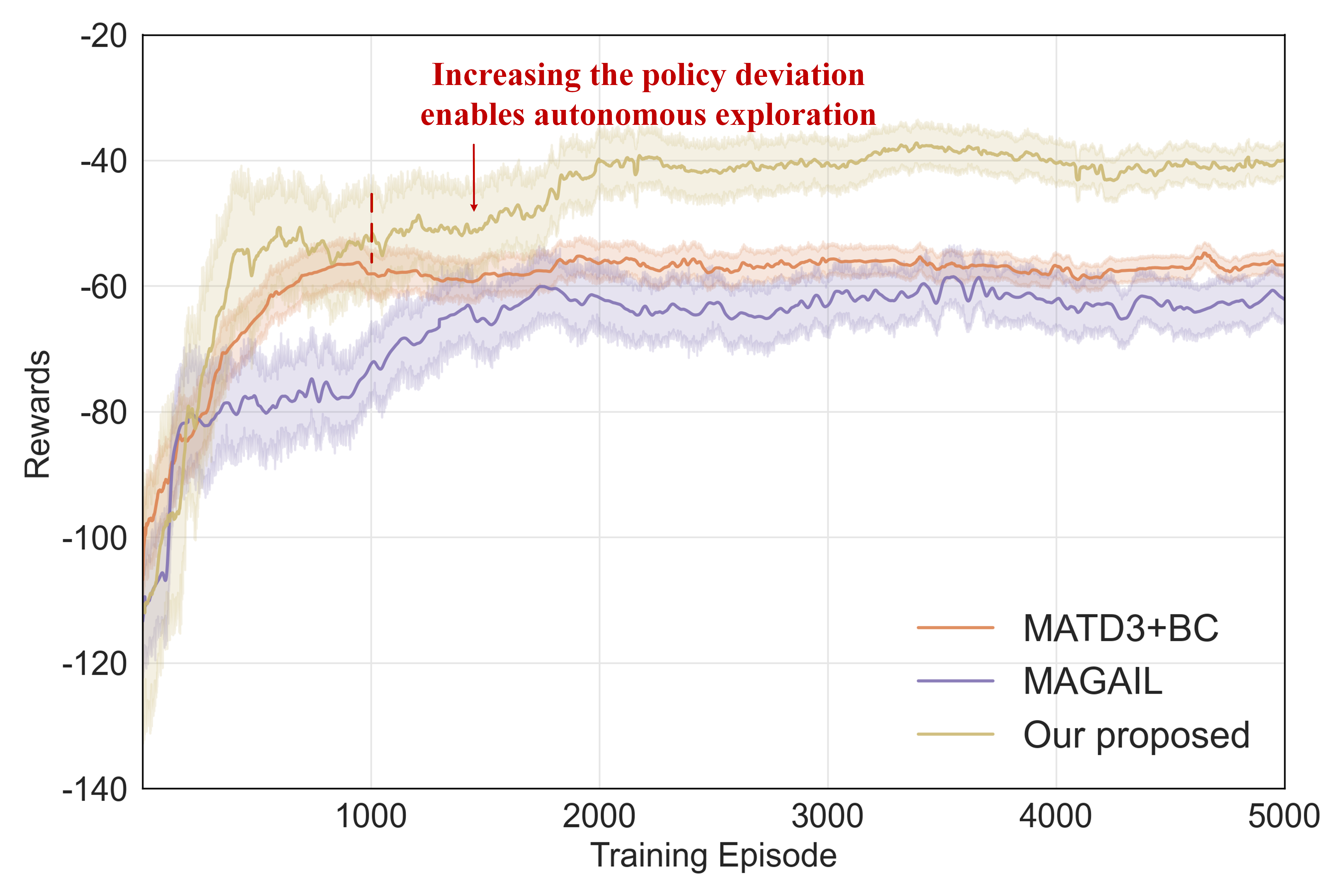}
	\caption{Training results of baseline algorithm under Qwen-3-8B strategies}
	\label{fig:explore8B}
\end{figure}

\hl{As shown in Fig. } \ref{fig:explore8B}, \hl{conventional imitation learning (e.g., BC, GAIL) excessively constrains policies, causing agents to inherit expert bias and converge to inferior local optimal. In contrast, our method gradually relaxes the imitation constraint by increasing policy deviation. This expands the exploration space and reintroduces reward signals, enabling agents to escape LLM expert-induced suboptimal equilibrium. The early-phase alignment provides stable initialization, while later-phase self-exploration compensates for expert inaccuracies. This two-phase process effectively balances expert knowledge exploitation with autonomous exploration. Conversely, BC- and GAIL-based methods enforce strict adherence throughout, rendering them unable to recover from suboptimal expert guidance.}

\hl{To assess the scalability of Our proposed-MH algorithm under large-scale prosumer participation, further experiments are conducted on the IEEE 141-bus distribution network with 100 prosumers, the prosumer details are reported in TABLE} \ref{tab:heterogeneous-prosumers-100}.

\begin{table}[htbp]
	\centering
	\caption{Personalized configurations for 100 prosumers on IEEE-141 system}
	\adjustbox{max width=\linewidth}{
		\begin{tabular}{@{}cl*{5}{c}@{}}
			\toprule
			Node & CDG & WT & PV & BESS & CL & Prosumer Type \\
			\midrule
			10, 12, 18, 24, 28, 33 & \multirow{2}{*}{\checkmark} & \multirow{2}{*}{--} & \multirow{2}{*}{\checkmark} & \multirow{2}{*}{\checkmark} & \multirow{2}{*}{\checkmark} & \multirow{2}{*}{Commercial} \\
			45, 48, 61, 78, 94, 102 & & & & & & \\
			
			15, 21, 30, 37, 52, 59 & \multirow{2}{*}{--} & \multirow{2}{*}{\checkmark} & \multirow{2}{*}{\checkmark} & \multirow{2}{*}{\checkmark} & \multirow{2}{*}{\checkmark} & \multirow{2}{*}{Rural} \\
			71, 84, 96, 109, 123, 130 & & & & & & \\
			
			17, 26, 36, 44, 54, 67 & \multirow{2}{*}{\checkmark} & \multirow{2}{*}{\checkmark} & \multirow{2}{*}{--} & \multirow{2}{*}{--} & \multirow{2}{*}{\checkmark} & \multirow{2}{*}{Industrial} \\
			75, 88, 95, 111, 124, 133 & & & & & & \\
			
			19, 29, 39, 47, 57, 62 & \multirow{2}{*}{--} & \multirow{2}{*}{--} & \multirow{2}{*}{\checkmark} & \multirow{2}{*}{\checkmark} & \multirow{2}{*}{\checkmark} & \multirow{2}{*}{Residential} \\
			76, 86, 99, 106, 121, 138 & & & & & & \\
			
			22, 31, 41, 50, 60, 74 & \multirow{2}{*}{\checkmark} & \multirow{2}{*}{\checkmark} & \multirow{2}{*}{\checkmark} & \multirow{2}{*}{\checkmark} & \multirow{2}{*}{\checkmark} & \multirow{2}{*}{Energy Hub} \\
			83, 100, 112, 116, 128, 134 & & & & & & \\
			
			6, 13, 23, 34, 46, 58, 68 & \multirow{2}{*}{--} & \multirow{2}{*}{\checkmark} & \multirow{2}{*}{\checkmark} & \multirow{2}{*}{\checkmark} & \multirow{2}{*}{--} & \multirow{2}{*}{Renewable} \\
			82, 97, 108, 114, 126, 135, 141 & & & & & & \\
			
			9, 16, 27, 35, 49, 57, 70 & \multirow{2}{*}{--} & \multirow{2}{*}{--} & \multirow{2}{*}{\checkmark} & \multirow{2}{*}{\checkmark} & \multirow{2}{*}{--} & \multirow{2}{*}{Storage-centric} \\
			81, 91, 101, 113, 125, 139 & & & & & & \\
			
			38, 56, 63, 72, 80, 90, 103 & \multirow{2}{*}{\checkmark} & \multirow{2}{*}{--} & 
			\multirow{2}{*}{--} & 
			\multirow{2}{*}{--} & \multirow{2}{*}{\checkmark} & \multirow{2}{*}{Data Center} \\
			110, 118, 129, 131, 136, 140 & & & & & & \\
			\bottomrule
		\end{tabular}
	}
	\label{tab:heterogeneous-prosumers-100}
\end{table}

\begin{figure*}[!t]
	\centering
	\subfloat[The smoothed curves of the average rewards\label{figure1-100}]{
		\includegraphics[width=2.25in]{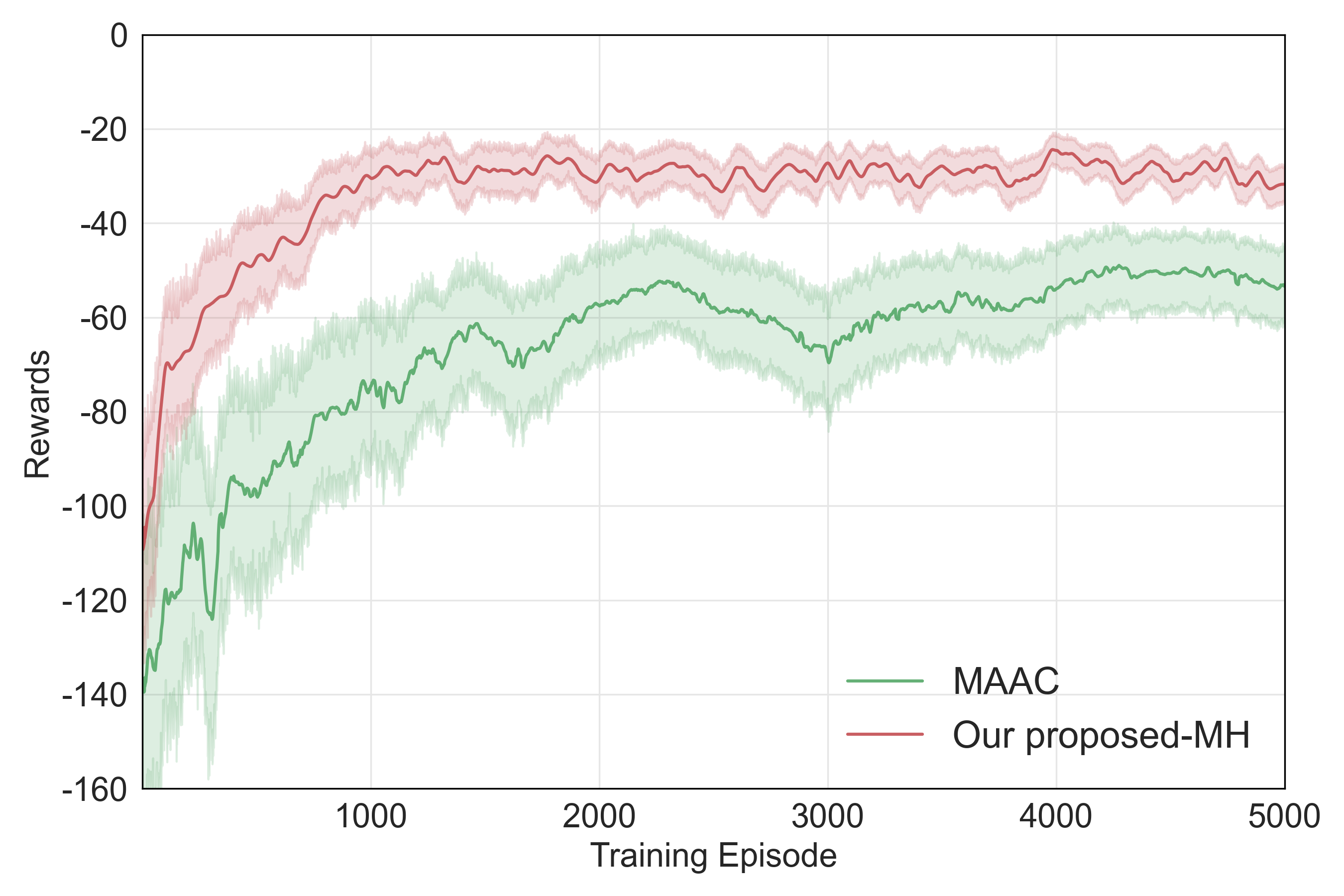}
	}
	\hfil
	\subfloat[The smoothed curves of the average operation cost\label{figure2-100}]{
		\includegraphics[width=2.25in]{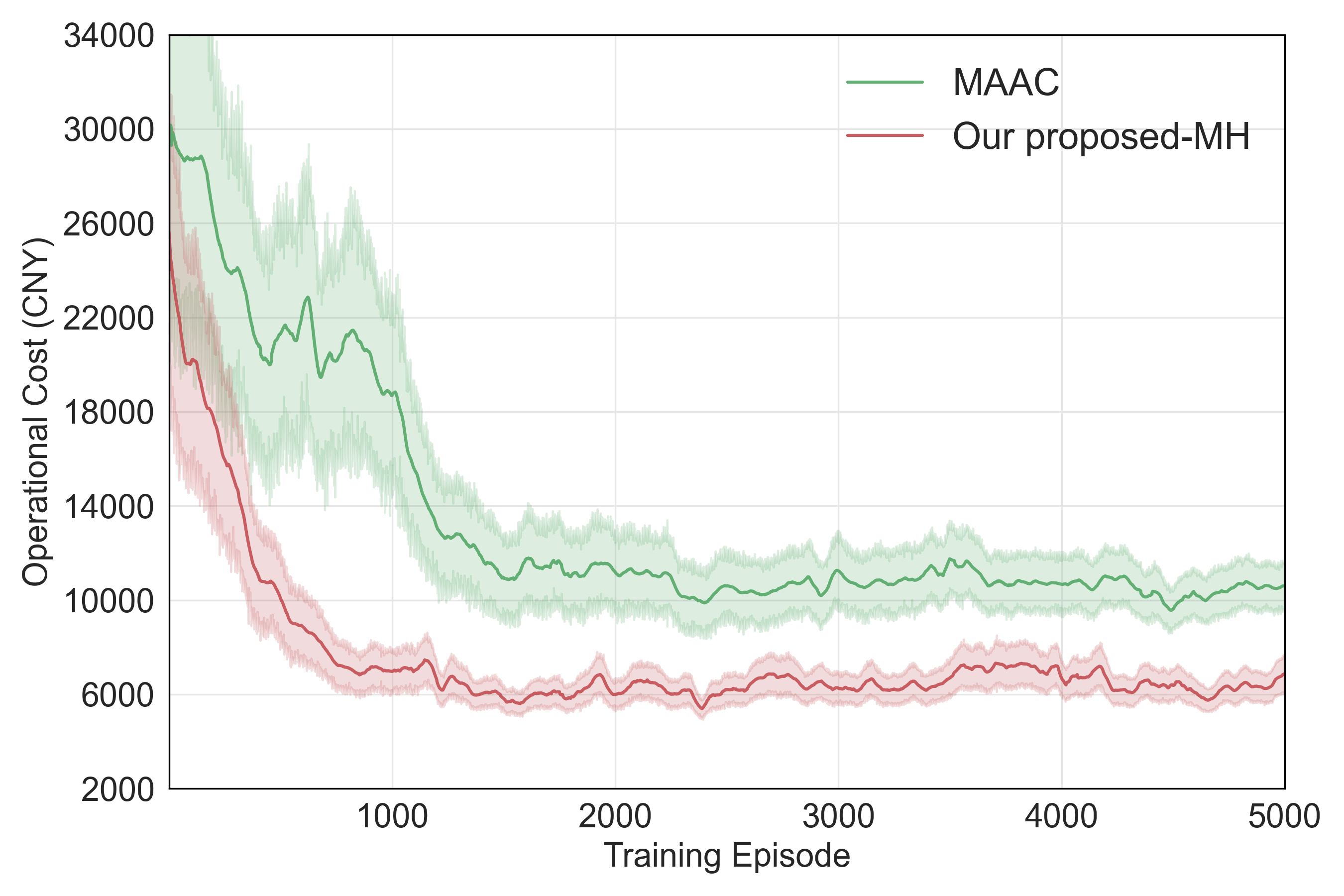}
	}
	\hfil
	\subfloat[The smoothed curves of the average voltage violation rate\label{figure3-100}]{
		\includegraphics[width=2.25in]{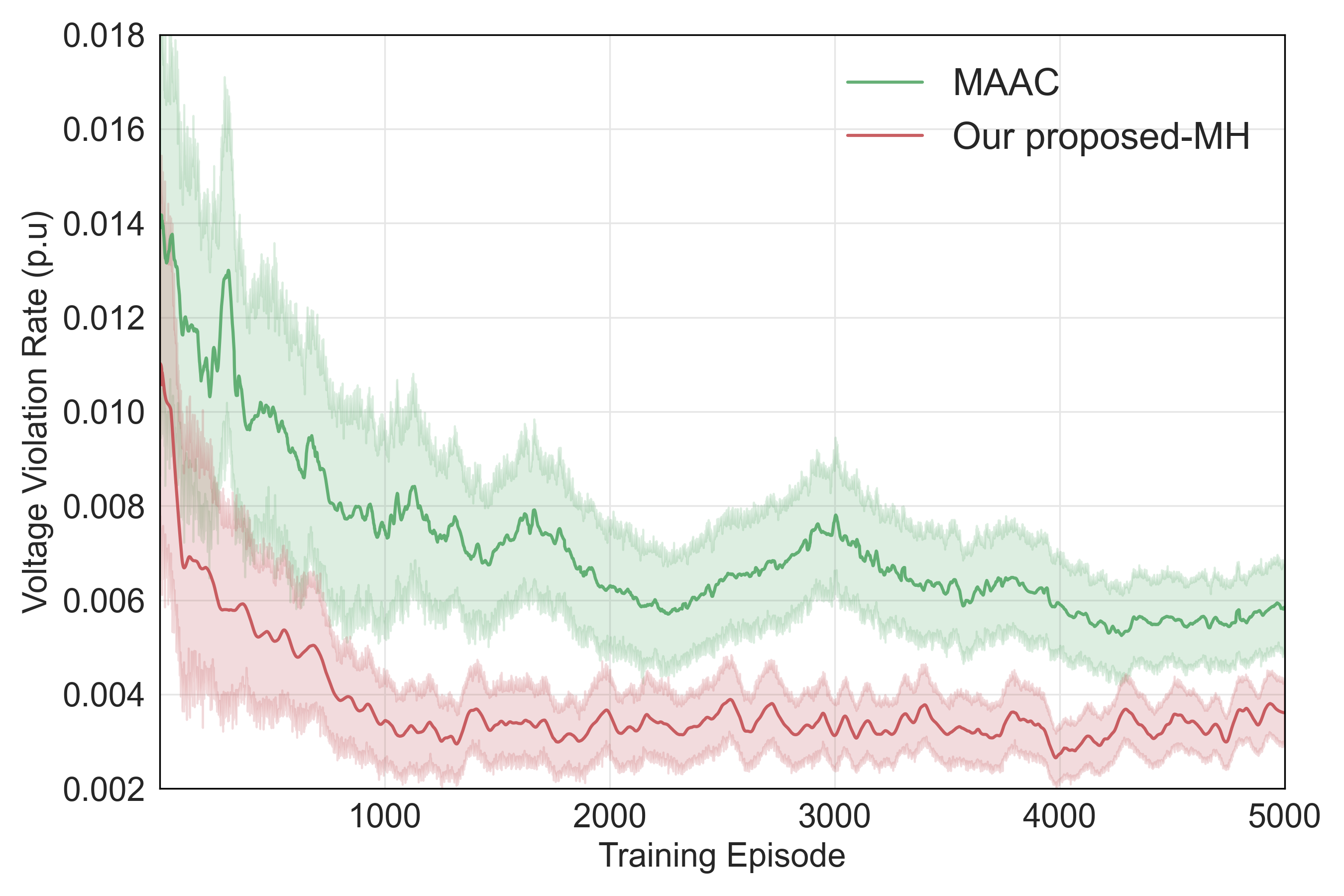}
	}
	\caption{Comparison chart of the performance of baseline algorithms with large-scale prosumers}
	\label{baselines-100}
\end{figure*}

\hl{Fig. } \ref{baselines-100} \hl{demonstrates the proposed method's scalability, with the LLM workflow replacing 14,874 lines of manual code. Conventional CTDE baselines (e.g., MATD3+BC, MAGAIL) fail in this setting as they require training 100 distinct critics with linearly growing input dimensions. In contrast, our proposed-MH algorithm addresses these limitations using a shared critic and Differential Attention, which ensures the concatenated representation ($Concat(e_i,x_i)$) remains fixed regardless of system scale. Together with agent-specific LLM guidance, this guarantees effectiveness in large-scale scenarios.}

\begin{figure}[ht]
\centering
\begin{subfigure}[b]{0.5\textwidth}
  \centering
  \includegraphics[width=0.95\linewidth]{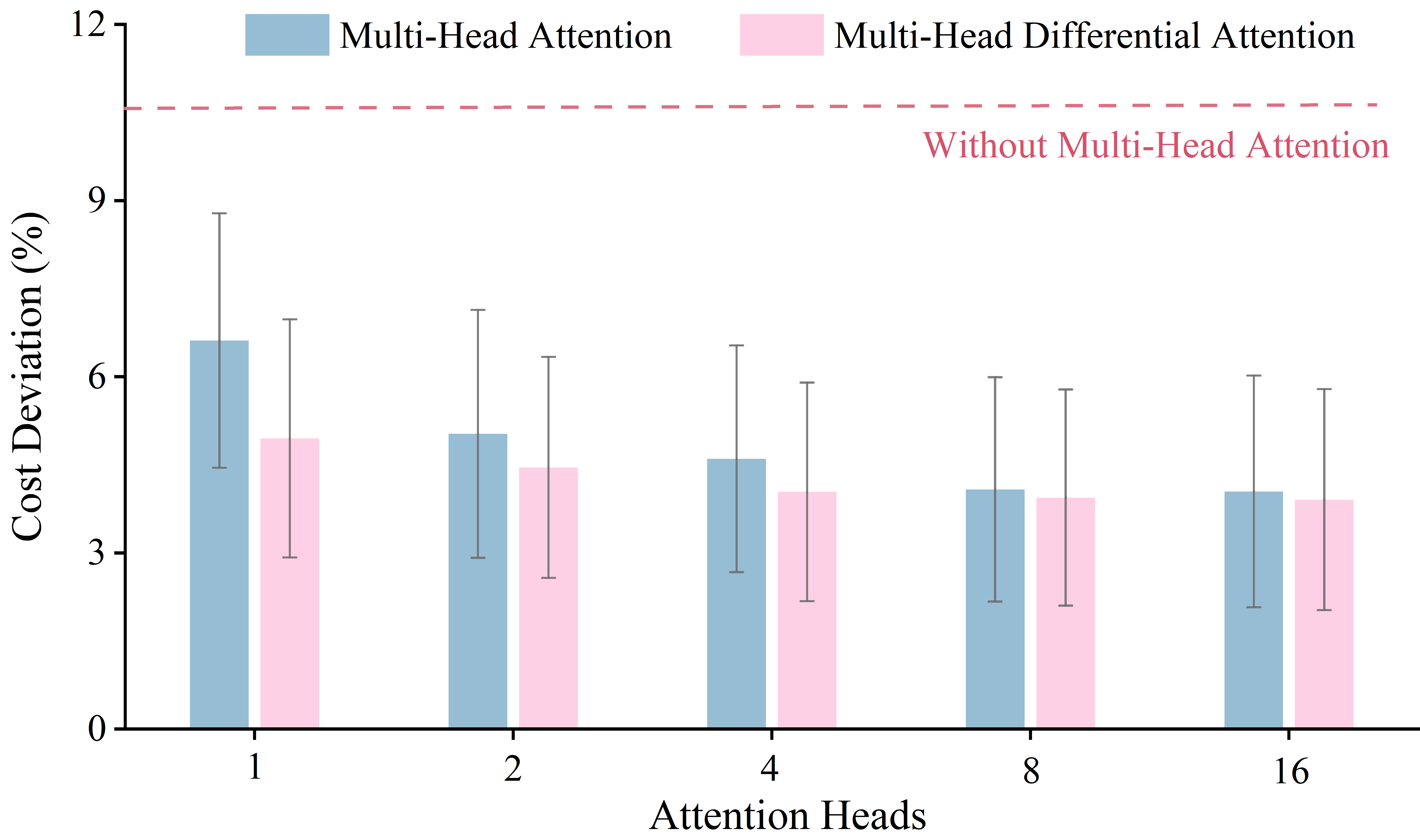}
\end{subfigure}
\caption{Ablation study on the number of attention heads in 20 agent prosumer}
\label{fig:head1}
\end{figure}

\hl{To evaluate the effectiveness of the proposed differential attention mechanism in multi-agent P2P energy trading, ablation studies were conducted, with results shown in Fig. }\ref{fig:head1}. \hl{The experimental results indicate that the number of attention heads substantially affects the cost deviation in the LLM expert. Increasing the number of heads from very few (e.g., 1–2) leads to a notable reduction in cost deviation, suggesting that introducing a small number of attention heads significantly enhances the model's ability to capture key agent interactions. However, further increasing the head count (e.g., from 8 to 16) yields only marginal improvements, indicating performance saturation likely due to redundancy in the captured features. Crucially, the differential attention mechanism outperforms standard attention in scheduling accuracy while preserving similar computational complexity. Its two sequential softmax operations incur roughly the same cost as a standard multi-head attention with twice the number of heads, thus delivering superior performance without computational overhead.}

\begin{table}[ht]
	\centering
	\caption{Ablation Study of Initial $\xi$}
	\label{tab:ablation_init}
	\small
	\setlength{\tabcolsep}{4pt}  
	\begin{tabular}{lcccccc}
		\toprule
		\multirow{2}{*}{\makecell{Initial \\ $\xi$}} & 
		\multicolumn{2}{c}{Episode 100} & 
		\multicolumn{2}{c}{Episode 1000} & 
		\multicolumn{2}{c}{Episode 5000} \\
		\cmidrule(lr){2-3} \cmidrule(lr){4-5} \cmidrule(lr){6-7}
		& Mean & Std.\ Dev. & Mean & Std.\ Dev. & Mean & Std.\ Dev. \\
		\midrule
		0.2 & -82.65 & 3.74 & -20.54 & 3.27 & -16.56 & 1.17\\
		0.5 & -85.23 & 4.13 & -20.20 & 3.31 & -17.30 & 1.22\\
		0.8 & -92.91 & 3.95 & -21.16 & 2.99 & -16.89 & 1.24\\
		\bottomrule
	\end{tabular}
\end{table}

\hl{An ablation study is conducted to investigate the impact of the initialization of the learnable vector $\xi$ in the Differential Attention mechanism. As shown in TABLE}  \ref{tab:ablation_init}, \hl{the choice of the initial value has only a minor effect on training performance during the early phase (within the first 100 training steps). After 1,000 training steps, the initialization of $\xi$ has a negligible influence on both the average training reward and its variance. These results indicate that $\xi$ can consistently adapt during training and gradually overcome the bias introduced by its initial value, demonstrating the robustness of the proposed Differential Attention mechanism with respect to parameter initialization.}

\hl{It is important to note that existing RL methods typically separate the training phase from real-world deployment, utilizing simulation environments rather than interacting directly with the real-world power system}\cite{11074719}. \hl{In this simulation-based training phase, traditional optimization methods are feasible as global information is accessible. However, they are often infeasible for real-time deployment due to the lack of accurate forecast parameters} \cite{9424985}.\hl{ Consequently, the LLM-generated optimization models are employed strictly to accelerate MARL training convergence. The trained agents are capable of autonomously executing energy trading tasks based solely on local observations, adaptively handling real-time uncertainties without further LLM intervention.}

{It should be noted that, like most RL approaches, the proposed framework exhibits limited generalization capability and requires retraining} \cite{pmlr-v97-cobbe19a} {whenever the environment undergoes substantial changes, such as major topology reconfiguration or the introduction of entirely new market mechanisms. Despite this inherent limitation of RL, the proposed LLM-MARL framework significantly outperforms conventional MARL methods in both convergence speed and final performance. To ensure reliability, all LLM generated strategies undergo DSO-based security verification, which guarantees their feasibility. Furthermore, the differential attention–based critic effectively mitigates performance degradation caused by increased agent interactions or system scale-up. As a result, the framework greatly reduces reliance on human expert guidance and significantly lowers the overall cost associated with full retraining in practical deployments.}

\section{CONCLUSION}
This paper addresses the collaborative decision-making challenges among multiple prosumers in local real-time P2P electricity markets by proposing a framework that integrates LLM expert guidance with MARL. This approach effectively overcomes limitations inherent in traditional optimization methods–particularly their inability to achieve real-time decision-making–as well as limitations of MARL without LLM guidance, particularly high manual labor costs associated with human expert involvement. The framework innovatively employs LLMs as experts to generate personalized strategies for guiding MARL training, combined with the Wasserstein metric and an enhanced Critic network, achieving deep integration of expert knowledge and agent learning. This significantly reduces manual costs while enhancing policy optimization performance.

Experimental validation demonstrates the method's superior performance. In model compatibility tests, the proposed framework exhibits universality across mainstream LLMs, with the Claude-4-Opus model achieving 100\% pass rate and 99.93\% accuracy in an expert workflow task, effectively substituting human experts. In the modified IEEE 141-bus distribution network, the proposed method achieves a remarkably low average operational cost of 4840.61 CNY and a voltage violation rate of only 1.06×$10^{-3}$ p.u., significantly outperforming conventional baseline methods.

\hl{Computationally, the framework's overhead is concentrated in the training stage, while real-time operation remains highly efficient using lightweight networks. The highest average cost of the LLM expert workflow is less than 6.27 CNY. Retraining is infrequent, triggered only by structural changes like topology modification or market adjustments, rather than routine fluctuations. Crucially, the training duration is negligible compared to the long intervals of these structural evolutions. Furthermore, the system supports efficient continual learning and knowledge transfer; the shared critic preserves learned network constraints, and the modular code repository enables direct reuse, significantly reducing redundancy and ensuring scalability.}

Future work will extend to larger-scale prosumer groups, expand external expert knowledge repositories, and explore the adaptability of diverse LLM workflow architectures in complex scenarios to strengthen the framework's practical value.





\ifCLASSOPTIONcaptionsoff
  \newpage
\fi

\end{document}